\documentclass[%
 reprint,
superscriptaddress,
groupedaddress,
 amsmath,amssymb,
 aps,
]{revtex4-1}
\usepackage{amsmath}
\numberwithin{equation}{section}
\renewcommand{\theequation}{\arabic{section}.\arabic{equation}}

\usepackage{dsfont} 

\usepackage{float}

\usepackage{graphicx}
\usepackage{dcolumn}
\usepackage{bm}
\usepackage{times}
\usepackage[colorlinks]{hyperref}
\hypersetup{
    citecolor=[rgb]{0,0,1},
    linkcolor=[rgb]{0,0,1},   
    urlcolor=[rgb]{0,0,0.9},
    linktocpage={true}}
\usepackage{wasysym}
\usepackage{mathtools} 

\usepackage[export]{adjustbox} 

\usepackage[mathscr]{euscript} 

\newcommand{\abs}[1]{\left\lvert#1\right\rvert} 

\usepackage{diagbox}

\usepackage{filecontents}

\begin{document}

\title{Non-diagonal anisotropic quantum Hall states}
\author{Pok Man Tam}
\author{Charles L. Kane}
\affiliation{Department of Physics and Astronomy, University of Pennsylvania, Philadelphia, PA 19104, USA}

\begin{abstract}
We propose a family of Abelian quantum Hall states termed the non-diagonal states, which arise at filling factors $\nu=p/2q$ for bosonic systems and $\nu=p/(p+2q)$ for fermionic systems, with $p$ and $q$ being two coprime integers. Non-diagonal quantum Hall states are constructed in a coupled wire model, which shows an intimate relation to the non-diagonal conformal field theory and has a constrained pattern of motion for bulk quasiparticles, featuring a non-trivial interplay between charge symmetry and translation symmetry. The non-diagonal state is established as a distinctive symmetry-enriched topological order. Aside from the usual $U(1)$ charge sector, there is an additional symmetry-enriched neutral sector described by the quantum double model $\mathcal{D}(\mathbb{Z}_p)$, which relies on the presence of both the $U(1)$ charge symmetry and the $\mathbb{Z}$ translation symmetry of the wire model. Translation symmetry distinguishes non-diagonal states from Laughlin states, in a way similar to how it distinguishes weak topological insulators from trivial band insulators. Moreover, the translation symmetry in non-diagonal states can be associated to the $\mathbf{e}\leftrightarrow\mathbf{m}$ anyonic symmetry in $\mathcal{D}(\mathbb{Z}_p)$, implying the role of dislocations as two-fold twist-defects. The boundary theory of non-diagonal states is derived microscopically. For the edge perpendicular to the direction of wires, the effective Hamiltonian has two components: a chiral Luttinger liquid and a generalized $p$-state clock model. Importantly, translation symmetry in the bulk is realized as self-duality on the edge. The symmetric edge is thus either gapless or gapped with spontaneously broken symmetry. For $p=2,3$, the respective electron tunneling exponents are predicted for experimental probes.
\end{abstract}

\maketitle

\tableofcontents

\section{\label{sec1}Introduction}
The discoveries of integer and fractional quantum Hall effects have initiated a revolution in the study of condensed matter that highlights the interplay between topology and physics \cite{Klitzing80, Tsui82, Laughlin1983}. In particular, fractional quantum Hall (FQH) states are characterized by topological order, which transcends Landau's paradigm of spontaneous symmetry-breaking \cite{Wen2004QFT}. Topologically ordered states host anyons, which are point-like quasiparticle-excitations that obey neither the bosonic nor fermionic exchange statistics. Rather, anyons have fractional statistics, and depending on whether they have a single or multiple fusion channels, they are classified as Abelian or non-Abelian anyons respectively. On practical grounds, there have been many research efforts investigating non-Abelian topological phases since certain types of non-Abelian anyons, such as the Fibonacci anyon and the Ising anyon, can support universal quantum computation \cite{Freedman2002QC, Freedman2003TQC, Kitaev2003TQC, Nayak2008TQCReview}. By braiding Fibonacci anyons, all possible unitary gates can be implemented with intrinsic fault-tolerance \cite{Simon2005Fibonacci,Simon2007Fibonacci}. By braiding Ising anyons, supplemented with a single-qubit phase gate and a two-qubit measurement gate, universal quantum computation can also be realized given rather mild error-correcting protocols \cite{Bravyi2006Ising, Freedman2006Ising}. There are also proposals of using Abelian anyons for quantum computations, with some but not all of the robustness as provided by non-Abelian anyons \cite{Lloyd2002AbelianTQC, Pachos2006AbelianTQC, QCToricCode2007}.   

On fundamental grounds, FQH states are prototypical platforms exhibiting the bulk-boundary correspondence. The bulk of a FQH state is gapped and characterized by a 2+1D topological quantum field theory, in which gauge fluxes are attached to quasiparticles \cite{Witten1989TQFT, Zhang1989EFT, Wen1990EFT, Wen1990EFT2, Read1990EFT, Fradkin1991CS, WenZee1992CS}. The boundary theory is gapless and described by a closely related 1+1D conformal field theory (CFT), where the primary fields are associated with the quasiparticle types \cite{Wen1991edge, MR1991edge, Wen1992edge, ReadRezayi1999, Hansson2017CFTReview}. The simplest example is the Laughlin state at filling $\nu=1/k$, whose bulk is described by the $U(1)_k$ Chern-Simon theory and the edge theory is the circle CFT (a free boson compactified on a circle) with radius $R=1/\sqrt{k}$ and chiral central charge $c=1$.


The bulk-edge correspondence of a topological phase is particularly manifest in a coupled-wire construction \cite{CWC02, CWC14}. There are two central ingredients in a coupled-wire construction. First, each wire is a one-dimensional Luttinger liquid consisting of two decoupled chiral and anti-chiral gapless modes. These decoupled modes are selected by tuning the intra-wire forward-scattering appropriately. Second, an array of wires interact together such that the chiral mode on one wire is coupled to the anti-chiral mode on the next wire, leading to a two-dimensional bulk that is completely gapped. At the end, a pair of gapless chiral modes remain, which are separated by the gapped bulk and localized at the boundary. This is conceptually similar to what happens in the non-trivial phase of the Su-Schrieffer-Heeger model \cite{SSH1980}, where inter-site couplings are engineered such that a physical electron can ``split" into two halves, each residing at a domain wall. Following this spirit, the coupled-wire construction has been applied to study various Abelian and non-Abelian FQH states \cite{CWC02, CWC14, CWC17, Fuji2017CWC, CWC18, Yukihisa2019, CWC20}, quantum spin liquid states \cite{CSL2015, Debanjan2016Z2QSL, CSL2017}, as well as higher-dimensional topological phases \cite{3dCWC2016, Fuji2019coupled-layers}.

As an exactly solvable model, the wire construction also allows one to study the motion of quasiparticles explicitly. Quasiparticles correspond to kink-excitations defined on a link between two wires, and they can move around by acting local operators on individual wires. Thus, the theory of a single wire, which is a full non-chiral CFT, contains important information about the scattering pattern of quasiparticles. The way that chiral and anti-chiral sectors are sewed together defines the allowed physical operators that can act on a single wire, and subsequently defines the allowed operators that scatter quasiparticles in the wire model \citep{CWC14}. When the wire is described by a diagonal CFT, which is the case for a Laughlin state, a local operator is given by a diagonal combination of chiral and anti-chiral fields, implying that all quasiparticles can be scattered across a single wire, having essentially unconstrained motion in the bulk. The purpose of this work is to introduce a simple but non-trivial twist to the Laughlin state, where a single wire is described by a \textit{non-diagonal} circle CFT. The resulting Abelian FQH state, termed as the non-diagonal state, has an interesting and constrained pattern of quasiparticle motion, which have significant physical consequences that we investigate below. 


In this paper, we propose a family of non-diagonal QH states using a coupled wire construction. The bosonic non-diagonal states arise at filling fraction $\nu=p/2q$, and the fermionic non-diagonal states arise at $\nu= p/(p+2q)$, with $p$ and $q$ being two coprime integers. For $p=1$, our construction produces the well-known Laughlin states, but there are interesting physics to be unveiled for $p>1$. Importantly, the non-diagonal QH states serve to highlight not only the interplay between topology and physics, but also the interplay between symmetry and topological order. As we will see, in the absence of either the $U(1)$ charge symmetry or the $\mathbb{Z}$ translation symmetry of the wire model, a non-diagonal state cannot be distinguished from a Laughlin state of charge-$pe$ particles, which is also known as a strongly-clustered state \cite{CWC17, CWC18}. Non-diagonal states thus share the same intrinsic topological order with strongly-clustered states. With charge conservation alone, they are both characterized by a $U(1)$ charge sector and possess a chiral Luttinger liquid on the boundary. However, in the presence of both charge symmetry and discrete translation symmetry, the non-diagonal states possess a distinct \textit{symmetry-enriched} topological (SET) order \cite{Ran2013SET, Hung2013SET, Lu2016SET, Cheng2016translationSET, Chen2017SET}. There is an additional symmetry-enriched neutral sector characterized by the quantum double model $\mathcal{D}(\mathbb{Z}_p)$, which has a $\mathbb{Z}_p$ topological order. 

Earlier on, the $\mathbb{Z}_p$ topological order has been realized in spin/rotor models such as Kitaev's toric code \cite{Kitaev2003TQC, Kitaev2006exactly, Bombin2010AS} and Wen's plaquette model \cite{You2012plaquette, You2013plaquette}, and in this work, the non-diagonal QH state is introduced as a platform for realizing $\mathcal{D}(\mathbb{Z}_p)$ in an electronic setting. Similar to the lattice models with spins, the translation symmetry in the coupled wire model also plays the role of the $\mathbf{e}\leftrightarrow\mathbf{m}$ anyonic symmetry of the $\mathbb{Z}_p$ topological order. In fact, the constrained motion of quasiparticles in the non-diagonal states distinguishes quasiparticles excited on the even links from those on the odd links, and as we will see, excitations on even and odd links can be respectively associated to the $\mathbf{e}$-type and $\mathbf{m}$-type anyons in $\mathcal{D}(\mathbb{Z}_p)$. Translation by a wire then interchanges even and odd links, thus acting as an anyon-relabelling transformation. A similar mechanism has been featured in the work by Hong and Fu \cite{Hong2017TCIsurface}, where a fermionic $\mathbb{Z}_4$ topological order is realized on the surface of topological crystalline insulators. Consequently, a dislocation in our wire model, which corresponds to a sudden termination of a wire inside the bulk, acts as a two-fold twist-defect that interchanges $\mathbf{e}$ and $\mathbf{m}$ anyons \cite{Bombin2010AS}. We expect the proposed non-diagonal QH states to be a concrete test bed for the general theory of anyonic symmetry \cite{Teo2015twistliquid, Teo2016AS, Barkeshli2019AS}. Moreover, it maybe interesting for future studies to explore the possibility of experimentally realizing the non-diagonal states in twisted-bilayer materials, as an array of quasi-one-dimensional subsystems could be engineered there, with the required translation symmetry built-in \cite{Kennes20201dflatbands}. 


The rest of the paper is organized as follows. In Sec. \ref{sec2}, we present a detailed study of the coupled-wire construction for the non-diagonal quantum Hall states. The existence of non-diagonal states is established in Sec. \ref{sec2.1}, followed by a discussion that explains the relation to non-diagonal CFTs in Sec. \ref{sec2.2}. In Sec. \ref{sec2.3}-\ref{sec2.4}, we analyze the scattering pattern of quasiparticles for both bosonic and fermionic non-diagonal states by explicitly constructing physical operators that move the quasiparticles around. The analysis allows us to appreciate the importance of charge conservation in constraining the motion of quasiparticles. We also provide an analogue of non-diagonal QH states in the context of weak topological superconductors, which contains a similarly constrained motion of vortex excitations that can be understood using the fractional Josephson effect. In Sec. \ref{sec3}, we study the non-diagonal states from the ``symmetry-enriched" perspective. We first establish a bulk neutral sector described by the $\mathcal{D}(\mathbb{Z}_p)$ quantum double model with a calculation of braiding statistics (Sec. \ref{sec3.1}), and discuss several concrete examples of non-diagonal states (Sec. \ref{sec3.2}). We then clarify the importance of the $U(1)$ charge symmetry and the $\mathbb{Z}$ translation symmetry in giving rise to the $\mathcal{D}(\mathbb{Z}_p)$ neutral sector (Sec. \ref{sec3.3}), thus establishing the non-diagonal state as possessing an SET order distinctive from the strongly-clustered Laughlin state. In Sec. \ref{sec4}, we present a detailed study of the boundary theory. We focus on the edge running perpendicular to the direction of wire, which is capable of realizing the translation symmetry. The corresponding effective Hamiltonian is found to consist of a chiral Luttinger liquid (Sec. \ref{sec4.1}) and a generalized $p$-state clock model (Sec.\ref{sec4.2}). Our discussion here corroborates and elaborates on some earlier works on critical parafermion chains \cite{criticalparafermionchain} and twist-defect chains \cite{twistdefectchain}. Translation symmetry in the bulk is realized as self-duality on the edge. The symmetric edge for $p=2,3$ is completely gapless, allowing for tunneling of a single electron into the edge, say from a Fermi liquid. Tunneling exponents are calculated in Sec. \ref{sec4.3}, which hopefully serve as an experimental probe for the non-diagonal states. Some complexities for the edge structure for $p\geq 4$ are addressed in Sec. \ref{sec4.4}, which supplement the results in earlier works \cite{twistdefectchain, criticalparafermionchain}. We conclude with outlook in Sec. \ref{sec5}. 

\section{\label{sec2}Wire model}
First let us describe a coupled wire construction for a sequence of $c=1$ Abelian FQH states, which bear an intimate relation to the non-diagonal series of circle CFT. They would thus be referred to as the ``non-diagonal" states. Though our construction may look similar to earlier formulations of the coupled-wire model \citep{CWC02, CWC14}, it sets the stage for exploring a subtlety that has been overlooked. One important consequence lies in the pattern of quasiparticle scattering, which will be investigated explicitly in our wire model. Bulk quasiparticles in general have a constrained motion, which reflects a non-trivial interplay between charge conservation and translation symmetry in the non-diagonal states. The discussion in this section sets the stage for identifying a hidden neutral sector as the $\mathbb{Z}_p$ toric code, as to be explained in Sec. \ref{sec3}. 

For the simplicity of exposition, the coupled wire construction is done for \textit{bosonic} electrons in this section. The fermionic non-diagonal states can be constructed in a similar manner, with the details presented in Appendix \ref{secappendixa}. While the relation to non-diagonal CFT is more transparent for the bosonic case, both bosonic and fermionic non-diagonal states share similar properties which we discuss in Sec. \ref{sec2.4}. 


\subsection{\label{sec2.1}Inter-wire coupling}
\begin{figure}[b!]
   \includegraphics[width=8cm,height=6.5cm ]{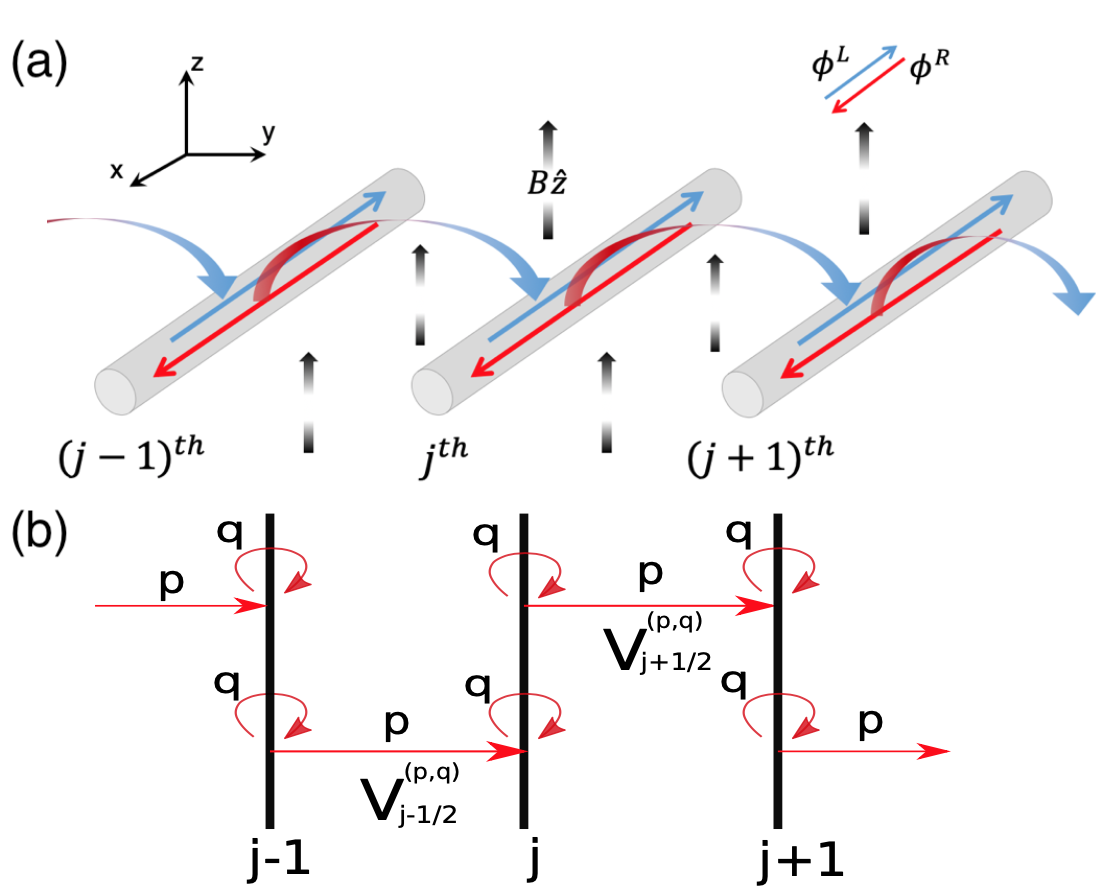}\centering
  \caption{\small{(a) Schematic of the coupled wire model. (b) Pictorial representation of the inter-wire coupling. }}
  \label{setup}
\end{figure}

We consider a two-dimensional system consisting of an array of $M$ one-dimensional wires of bosons, as depicted in Fig. \ref{setup}(a). A perpendicular magnetic field is applied, and the one-dimensional (along wire) flux density is denoted as $b$. The low energy description of this system is provided by ``bosonizing the bosons", such that each wire is characterized by two slowly varying bosonic fields: the phase variable $\varphi(x)$ and the density variable $\theta(x)$ \citep{giamarchi2003, gogolin2004}. They form a conjugate pair, satisfying the following canonical commutation relation:
\begin{equation}
[\partial_x\theta_j(x), \varphi_{j'}(x')] = i\pi\delta_{jj'}\delta(x-x'),
\end{equation}
with $j,j'=1, 2, ..., M$ labeling the wires and $x,x'$ being the coordinate along the quantum wires.  The operator that annihilates an (bosonic) electron on the $j$-th wire is then expressed as
\begin{equation}\label{phi}
\psi_j(x) \propto e^{i\varphi_j(x)}.
\end{equation}
The operators associated to density fluctuation are expressed as
\begin{equation}\label{theta}
\rho^n_j \propto e^{i2n(\pi \bar{\rho}x+\theta_j(x))},
\end{equation}
with $n\in \mathbb{Z}$ and $\bar{\rho}$ being the 1d average density. This describes an intra-wire back-scattering at wavevector $k\sim 2n\pi \bar{\rho}$.

A crucial ingredient of a wire construction is the inter-wire coupling, which involves tunneling of electrons across neighboring wires. Ultimately, it determines how the bulk is gapped to give rise to a quantum Hall state, as well as what kind of gapless edge is left at the boundary. For our purpose, the tunneling interaction is characterized by a pair of coprime integers $(p,q)$, which describes the inter-wire tunneling of $p$ electrons between two nearest wires, accompanied by intra-wire backscattering at wavevector $k\sim 2q\pi\bar{\rho}$. A pictorial representation of this tunneling interaction is provided in Fig. \ref{setup}(b). The inter-wire coupling term, defined for each link $\ell=j+1/2$ between wires $j$ and $j+1$, then takes the following expression:
\begin{equation}\label{tunnel1}
\begin{split}
\mathcal{V}^{(p,q)}_{\ell} &= (\psi_{j+1}^\dagger \psi_j e^{-ibx})^p \rho^q_{j+1}\rho^q_j + h.c.\\
&=e^{i(4\pi\bar{\rho}q-pb)x} e^{i\Theta_\ell}+h.c.\;,
\end{split}
\end{equation}
with the following link variable defined:
\begin{equation}
\Theta_{\ell} = p(\varphi_{j}-\varphi_{j+1})+2q(\theta_{j}+\theta_{j+1}).
\end{equation}
Notice that the Lorentz force provides an impulse to each tunneled electron, which is accounted for by the $e^{-ibx}$ factor attached above. The oscillatory factor in Eq. (\ref{tunnel1}) describes the net momentum of the tunneling operator $\mathcal{V}^{(p,q)}_{\ell}$. Demanding momentum conservation, we obtain the filling fraction for the bosonic FQH state under consideration,
\begin{equation}\label{filling}
\nu \equiv \frac{2\pi \bar{\rho}}{b} = \frac{p}{2q}.
\end{equation}

Next, we proceed to demonstrate how the inter-wire coupling gaps out the bulk, and then examine the gapless chiral modes left at the boundary. To that end, it is convenient to introduce a set of chiral bosonic fields,
\begin{subequations}\label{chiralboson}
\begin{align}
\phi^\text{R}_j&=p\varphi_j+2q\theta_j,\\
\phi^\text{L}_j&=p\varphi_j-2q\theta_j, 
\end{align}
\end{subequations}
which can be checked to have the following commutation relation:
\begin{equation}\label{chiralcommutation}
[\partial_x\phi^{\tilde{r}}_j(x),\phi^{\tilde{r}'}_{j'} (x')] = 4i\pi pq \tilde{r}\delta_{\tilde{r}\tilde{r}'}\delta_{jj'}\delta(x-x'),
\end{equation}
where $\tilde{r},\tilde{r}'=\text{R}/\text{L}=+1/-1$. From the Luttinger liquid theory \citep{giamarchi2003, gogolin2004}, it is known that with an appropriate intra-wire forward-scattering the Luttinger parameter can be adjusted such that the Hamiltonian of a single wire takes the following form,
\begin{equation}\label{singlewire}
\mathcal{H}_j = \frac{u}{2\pi}[(\partial_x\phi^\text{R}_j)^2+(\partial_x\phi^\text{L}_j)^2], 
\end{equation}
In this way, the designated chiral bosonic fields are decoupled in each wire. (Here $u$ is the speed of sound, which is not essential to our later discussions.) 
Subsequently, the complete Hamiltonian for the coupled wire model is
\begin{equation}\label{wireHamtot}
\mathcal{H}_{\rm tot}=\sum_{j=1}^M \mathcal{H}_j + \sum_{j=1}^{M-1}t_{j+\frac{1}{2}}\cos\Theta_{j+\frac{1}{2}}.
\end{equation}
The second term comes from the tunneling operator $\mathcal{V}^{(p,q)}_{\ell}$ for each link $\ell=j+1/2$, with $t_{\ell}$ characterizing the tunneling strength. The filling fraction is adjusted so that the oscillatory factor is canceled. Notice that in terms of chiral bosonic fields, the link variable can be expressed as
\begin{equation}\label{link1}
\Theta_{j+\frac{1}{2}} = \phi^\text{R}_j-\phi^\text{L}_{j+1}.
\end{equation}
Therefore, while the chiral modes are decoupled within individual wires, the tunneling operators are simply coupling chiral modes with opposite chirality from neighboring wires, as illustrated in Fig. \ref{setup}(a). This simple picture motivates us to analyze the interacting Hamiltonian in Eq. (\ref{wireHamtot}) by the following decomposition:
\begin{equation}
\mathcal{H}_{\rm tot} = \mathcal{H}_{\rm edge} +  \mathcal{H}_{\rm bulk},
\end{equation}
with the contribution from the boundary being
\begin{equation}\label{edgeHam}
\mathcal{H}_{\rm edge} = \frac{u}{2\pi}[(\partial_x \phi^\text{L}_1)^2+(\partial_x \phi^\text{R}_M)^2],
\end{equation}
and the contribution from the bulk being
\begin{equation}\label{bulkHam}
\begin{split}
\mathcal{H}_{\rm bulk} = \sum_{j=1}^{M-1}\{ \frac{u}{4\pi}[(\partial_x \Phi_{j+\frac{1}{2}})^2+(\partial_x\Theta_{j+\frac{1}{2}})^2] \\
 +\;t_{j+\frac{1}{2}}\cos\Theta_{j+\frac{1}{2}}\}.
\end{split}
\end{equation}
Here we have also introduced the conjugate link variable
\begin{equation}\label{link2}
\Phi_{j+\frac{1}{2}} = \phi^\text{R}_j+\phi^\text{L}_{j+1}
\end{equation}
which, together with the link variable defined earlier in Eq. (\ref{link1}), obey the following commutation relation:
\begin{equation}\label{linkcommutation}
[\partial_x\Theta_{\ell}(x) , \Phi_{\ell'} (x')] = 8i\pi p q \delta_{\ell \ell'} \delta(x-x').
\end{equation}

The bulk Hamiltonian $\mathcal{H}_{\rm bulk}$ can now be viewed as decoupled copies of sine-Gordon models, each for a link. In particular, when the interaction term $\cos\Theta_{\ell}$ flows to strong coupling, the link variables $\Theta_{\ell}$ in the bulk are all pinned at the bottom of the cosine potential, \text{i.e.} $\Theta_{\ell}=2\pi n\; (n\in \mathbb{Z})$, which leads to a gapped bulk. Such an exactly solvable limit of our interacting microscopic model can be attained when we include additional \textit{inter-wire scattering} of the form: $(\partial_x\phi^\text{R}_{j})(\partial_x\phi^\text{L}_{j+1})$. The net effect of inter-wire scattering can be absorbed into the Luttinger parameter $K$, so the bulk Hamiltonian is modified to:
\begin{equation}\label{bulkHam2}
\begin{split}
\mathcal{H}'_{\rm bulk} = \sum_{j=1}^{M-1}\{&\frac{u}{4\pi}[K(\partial_x \Phi_{j+\frac{1}{2}})^2+\frac{1}{K}(\partial_x\Theta_{j+\frac{1}{2}})^2]\\
&+\; t_{j+\frac{1}{2}}\cos\Theta_{j+\frac{1}{2}}\}.
\end{split}
\end{equation}
For $K< (pq)^{-1}$, the scaling dimension of the inter-wire tunneling is found to be $\Delta_t<2$ and thus the cosine potential is relevant. This gives an exactly solvable regime of our model, in which the bulk is known to be gapped. Up to this point, we have started from an interacting microscopic model and constructed a family of bosonic quantum Hall states at filling fraction $\nu=p/2q$. For $p=1$, these states are simply the Abelian Laughlin states which have been discussed before \citep{CWC14}. Indeed, even for generic values of $(p,q)$, the corresponding state is still Abelian. Thus the reader may wonder whether we can learn anything exciting here by studying the generic case with $p>1$. The answer is surprisingly affirmative, and has to do with how quasiparticles are scattered (\textit{i.e.} their allowed motion) in the wire model. 

In the coupled wire construction, quasiparticles appear at link $\ell$ when $\Theta_{\ell}$ has a kink where it jumps by $2\pi n $ ($n\in \mathbb{Z}$) \citep{CWC02, CWC14}. Away from the kink, the system is still in its ground state as suggested by Eq. (\ref{bulkHam2}), and around the kink there is an accumulation of charge $ne/2q$. Following Eq. (\ref{chiralboson}) and Eq. (\ref{link1}), a charge-$e/2q$ quasiparticle residing at link $j+1/2$  can be created by the following operator:
\begin{equation}\label{QPcreation}
(\Psi^{\text{R}/\text{L}}_{e/2q, j+\frac{1}{2}} (x))^\dagger= e^{-\frac{i}{2pq}\phi^{\text{R}/\text{L}}_{j/j+1}(x)}.
\end{equation}
This is not a local operator, as anticipated because quasiparticles cannot be created locally. On the other hand, scatterings of quasiparticles are expected to be represented by local operators. When acting on a single wire, it is clear from Eqs. (\ref{phi}) and (\ref{theta}) that local operators should take the form $e^{i(r\varphi-2s\theta)}$ ($r,s\in \mathbb{Z}$). One can then check that the minimal quasiparticle with charge $e/2q$ can be scattered across a \textit{single} wire only if the scattering operator
\begin{equation}\label{scattop}
\mathcal{O}_j= e^{\frac{i}{2pq}(\phi^\text{R}_j-\phi^\text{L}_j)} = e^{\frac{2i}{p}\theta_j}
\end{equation}
is local, which is true only when $p=1$. When $p>1$, a minimal quasiparticle needs to hop across \textit{two} wires in order to obey locality and preserve its charge, which we will explain in more detail. The motion of quasiparticle is thus constrained in general, and as we soon see, this is a defining feature of ``non-diagonal" states. We want to emphasize that the phase we have constructed here is different from the strongly-clustered phase studied before in Ref. \citep{CWC17,CWC18}. In the strongly-clustered phase, electrons are bound into charge-$pe$ clusters and form the Laughlin state at filling $\nu_{pe} = 1/2pq$ (equivalently the electronic filling is $\nu=p/2q$), which also has quasiparticles with minimal charge $e/2q$. However in that case, irrespective of what $p$ is, quasiparticles can always hop across a single wire. 

The constrained motion of quasiparticles also appears in fermionic non-diagonal states. Moreover, it also happens in a superconducting context. There, the contrast between the non-diagonal state and the strongly-clustered state can be understood in terms of fractional Josephson effect, by considering an array of 1d topological/trivial superconductors, with quasiparticles replaced by quantum vortices. We will elaborate more on this analogy, but before that, we want to introduce an intimate connection between the constructed FQH state to what is known as the \textit{non-diagonal} conformal field theory. The aforementioned pattern of bulk quasiparticle scattering can be understood from the perspective of the full non-chiral boundary CFT, or equivalently, the theory of a single wire.

\subsection{\label{sec2.2} Single wire: non-diagonal CFT}
While the bulk is gapped, it is manifest in the coupled wire model that there are gapless chiral modes left at the boundary, namely at the very first wire $(j=1)$ and the very last wire $(j=M)$. We have the edge Hamiltonian $\mathcal{H}_{\rm edge}$ written out in Eq. (\ref{edgeHam}), which resembles the Hamiltonian of a single wire in Eq. (\ref{singlewire}), except that the two chiral edge modes are separated by a gapped bulk. From Eq. (\ref{chiralcommutation}), it is clear that each edge is described by a chiral Luttinger liquid with Luttinger parameter $K=2pq$. The edge theory is equivalently known as the $U(1)_{2pq}$ chiral CFT, and it also describes the edge of the strongly-clustered state at filling $\nu_{pe}=1/2pq$. Hence, from the perspective of a single chiral sector at the boundary, the non-diagonal state is no different from the strongly-clustered state. To distinguish them, it is important to combine the chiral and anti-chiral sectors, and study the resulting non-chiral theory. This full non-chiral boundary theory, which is equivalent to the theory of a single wire, determines how quasipartilces can move in the bulk and be scattered from one edge to another. Here the non-chiral theory describes a boson compactified on a circle, and first let us determine its \textit{radius of compactification}.

The circle-compactification can be easily inferred when the bosonic fields $\varphi$ and $\theta$ were first introduced in Eqs. (\ref{phi}) and (\ref{theta}). They are defined to have the following shift-symmetries that leave the physical operators invariant:
\begin{equation}
\varphi \mapsto \varphi+2\pi, \;\;\;\theta \mapsto \theta +\pi.
\end{equation}
Consequently, the circle CFT with the Hamiltonian,
\begin{equation}
\mathcal{H} = \frac{u}{2\pi}[(\partial_x \varphi)^2+(\partial_x\theta)^2]
\end{equation}
is considered to have radius $R=1$ (when looking at the compactification of $\varphi$), or radius $R=1/2$ (when looking at the compactification of $\theta$). These two descriptions of radius are indeed equivalent, thanks to a duality between circle CFTs of radius $R$ and $1/2R$, which is known as the $T$-duality \citep{Ginsparg88,BYB}. 

Substituting Eq. (\ref{chiralboson}) into Eq. (\ref{singlewire}), the Hamiltonian describing a single wire can be expressed as
\begin{equation}
\mathcal{H}_j = \frac{\tilde{u}}{2\pi}[\frac{p}{2q}(\partial_x\varphi)^2+\frac{2q}{p}(\partial_x\theta)^2],
\end{equation}
with the speed of sound rescaled to $\tilde{u} = 4pqu$. Comparing with the circle CFT at radius $R=1$, the circle CFT corresponding to our single wire now has the following radius:
\begin{equation}
R=\sqrt{\frac{p}{2q}}.
\end{equation}
In this paper, we focus on the situation where $p$ and $q$ are coprime integers, as otherwise there exists two smaller coprime integers giving rise to the same radius for the edge theory, as well as same filling factor for the bulk. For $p=1$, the radius is $R=1/\sqrt{2q}$, which leads to the familiar circle CFT that is known to describe the gapless edges of the $\nu=1/2q$-filling Laughlin state, and as we will soon see, it is a \textit{diagonal} theory. Below, let us introduce the distinction between diagonal and non-diagonal CFTs for a boson compactified on a circle, by first studying their corresponding partition functions. 

As discussed by DiFrancesco \textit{et al.} \citep{BYB}, the modular-invariant partition function for a compact boson of rational radius $R= \sqrt{p/2q}$ can be expressed as follows \citep{Tduality},
\begin{equation}\label{partition}
Z(\sqrt{\frac{p}{2q}}) = \sum_{n=0}^{N-1}K_{
n}(\tau)\overline{K_{\omega n}(\tau)},
\end{equation}
with $\tau$ being the modular parameter and $K_n(\tau)$ the extended character which can be expressed as,
\begin{equation}\label{charac}
K_n(\tau) = \frac{1}{\eta(\tau)}\sum_{m\in \mathbb{Z}}\lambda^{(Nm+n)^2/2N},
\end{equation}
where $\eta(\tau)$ is the Dedekind eta function and $\lambda=e^{2i\pi\tau}$. In the above we have defined $N=2pq$, which counts the number of chiral primary fields. Modular-invariance requires the parameter $\omega$ in Eq. (\ref{partition}) to satisfy the following conditions:
\begin{subequations}\label{Bezout}
\begin{align}
qr_0-ps_0&=1,\\
qr_0+ps_0&= \omega \;\;\;\text{mod }N.
\end{align}
\end{subequations}
In the range $1\leq r_0 \leq p$, $1\leq s_0 \leq q-1$, the B\'ezout's lemma in number theory guarantees the unique existence of an integer solution $(r_0,s_0)$ to the first equation \cite{Bezout2009}, which subsequently defines $\omega$ in the second equation. For $p=1$, we have the solution $(r_0,s_0) = (1,q-1)$, which leads to $\omega = -1\;\text{mod }N$. From Eq. (\ref{charac}), it can be seen that the extended character obeys $K_{n} = K_{-n}$, and hence for $p=1$ we have,
\begin{equation}
Z(\frac{1}{\sqrt{2q}}) = \sum_{n=0}^{N-1} \abs{K_{n}}^2. 
\end{equation}
This defines the diagonal theory, in which the extended characters from the chiral and anti-chiral sectors are combined in a symmetric manner. On the contrary, for coprime integers $p,q > 1$, the partition function in Eq. (\ref{partition}) \textit{cannot} be expressed in the diagonal form, and the corresponding theory is known as non-diagonal \citep{BYB}. This explains why we would name the Abelian states constructed in Sec. \ref{sec2.1} as the ``non-diagonal" states for $p,q >1$. It is worth making a contrast with the strongly-$p$-clustered state at filling $\nu=p/2q$ \citep{CWC17,CWC18}, which also has a minimal quasiparticle of charge $e/2q$ but its edge theory has compactification radius $R=1/\sqrt{2pq}$, and is thus a ``diagonal" state.

In the coupled wire model, one can see an important physical consequence regarding the distinction between diagonal and non-diagonal theories \cite{offdiagonal}. Each extended chiral/anti-chiral character $K_{n}/\overline{K}_n$ is associated to a chiral/anti-chiral primary operator, $e^{\pm\frac{in}{2pq}\phi^{\text{R}/\text{L}}}$, while the expansion of the partition function in terms of extended characters, namely Eq. (\ref{partition}), suggests how a physical local operator can be constructed from a combination of chiral and anti-chiral sectors. For the $\nu=p/2q$ state constructed in the wire model, the theory of a single wire is non-diagonal for $p,q>1$, so the desired scattering operator $\mathcal{O}_j$ introduced in Eq. (\ref{scattop}), as a diagonal combination of chiral and anti-chiral primary operators, is \textit{not} an allowed local operator. Hence the minimal quasiparticle in a non-diagonal state cannot hop across just a single wire in the coupled wire model. But it can always hop across two, as we explain in the next subsection. This constraint leads to the distinction of two types of quasiparticles: one that lives on the even links and the other on the odd links. As we explain in Sec. \ref{sec3}, they are associated to the $\mathbf{e}$-type and $\mathbf{m}$-type anyons in the $\mathbb{Z}_p$ toric code, respectively. 

The states with $p>1$ and $q=1$ require special attention, as they are also ``non-diagonal". Admittedly, from the perspective of CFT, they have diagonal partition functions just like the diagonal states (with $p=1$ and $q>1$). In fact under $T$-duality, which interchanges $\varphi$ and $2\theta$, $p$ and $q$ are also interchanged, so this is expected. Nevertheless, $\varphi$ and $\theta$ have their respective physical meanings in the wire model. In particular, the electron operator $e^{i\varphi}$ carries charge-$e$ while the density operator $e^{i2\theta}$ is charge-neutral. As we see next, charge conservation (as a natural symmetry in a quantum Hall system) plays an important role in constraining the motion of quasiparticle. One thus should not use $T$-duality to disqualify the states with $p>1$ and $q=1$ as being ``non-diagonal". In fact, by analyzing the allowed quasiparticle scattering pattern, these states are constrained in just the same way as any other non-diagonal states.

\subsection{\label{sec2.3}Quasiparticle scattering}

\begin{figure}[t!]
   \includegraphics[width=6cm,height=4.5cm]{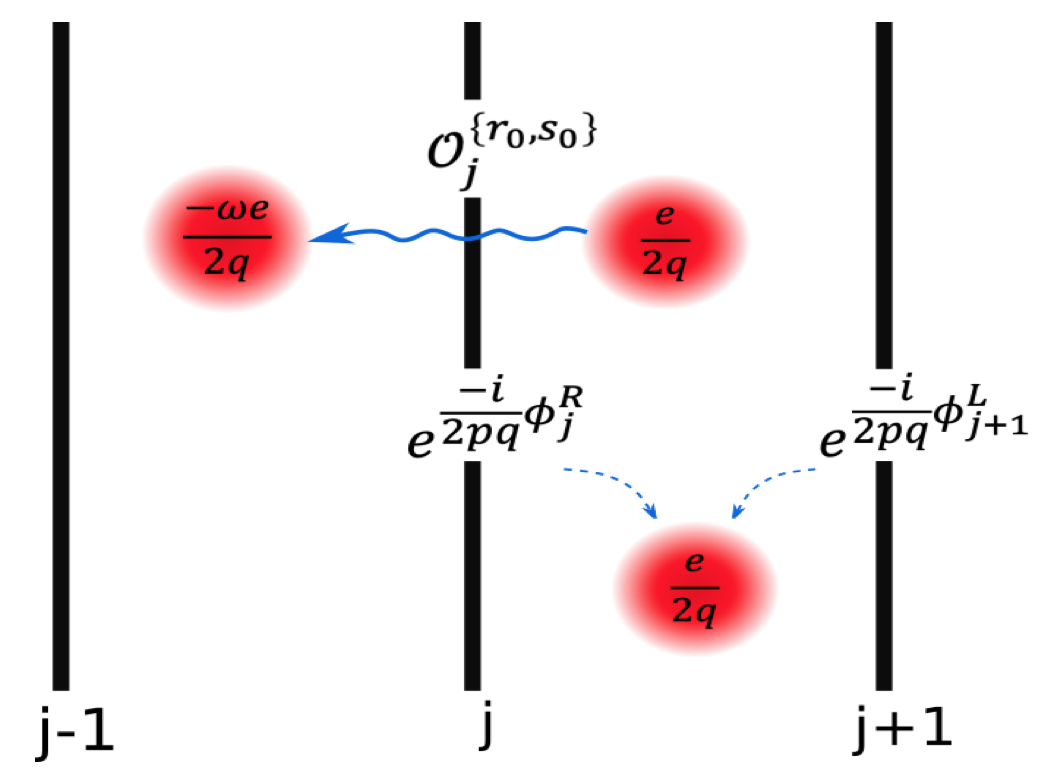}\centering
  \caption{\small{Schematic illustration of quasiparticle operators. Quasiparticles in the coupled wire model are excitations living on the links between consecutive wires. For the $\nu=p/2q$ Abelian quantum Hall states, the minimal non-trivial quasiparticle has charge $e/2q$. Such a quasiparticle on link $j+1/2$ can be created either by $e^{\frac{-i}{2pq}\phi^{\text{R}}_{j}}$ or $e^{\frac{-i}{2pq}\phi^{\text{L}}_{j+1}}$, which are non-local operators. Local operators defined in Eq. (\ref{localop}) scatter quasiparticles, and thus allow them to move in the bulk. For instance,  applying $\mathcal{O}^{\{r_0,s_0\}}_j$ would scatter a quasiparticle of charge $e/2q$ on link $j+1/2$ to a quasiparticle of charge $-\omega e/2q$ on link $j-1/2$, with $r_0, s_0$ and $\omega$ defined through Eq. (\ref{Bezout}).}}
  \label{qpoperators}
\end{figure}

Let us now discuss the allowed motion of bulk quasiparticles in more detail. Following the theory of Luttinger liquid and bosonization \citep{giamarchi2003, gogolin2004}, the smeared density on the $j$-th wire is
\begin{equation}
\rho_j = \frac{1}{\pi} \partial_x\theta _j = \frac{1}{4\pi q}(\partial_x\phi^\text{R}_j - \partial_x \phi^\text{L}_j),
\end{equation}
with the second equality following from the definition of chiral bosonic fields in Eq. (\ref{chiralboson}). Instead of assigning electric charge to the wires, one can equally well assign charge to the links and consider the following density operator for link $\ell = j+1/2$:
\begin{equation}\label{linkdensity}
\rho_{\ell} = \frac{1}{4\pi q} (\partial_x \phi^\text{R}_{j}-\partial_x \phi^\text{L}_{j+1}) = \frac{1}{4\pi q} \partial_x\Theta_\ell,
\end{equation}
with the second equality following from the definition of link variable $\Theta_\ell$ in Eq. (\ref{link1}). In our wire construction, a quantum Hall state is established when all link variables $\Theta_{\ell}$ are pinned at the bottom of the cosine potential (arising from inter-wire coupling), where they take values in $2\pi n$ ($n\in \mathbb{Z}$). Therefore, a minimal non-trivial quasiparticle excitation located on link $\ell$ would correspond to a $2\pi$-kink in $\Theta_{\ell}$. The density operator in Eq. (\ref{linkdensity}) suggests that this quasiparticle carries charge $e/2q$. The creation/annihilation operator for such a quasiparticle can be constructed by requiring it to create a $2\pi$-kink in $\Theta$. We have already introduced them ahead of time in Eq. (\ref{QPcreation}), where it is used to illustrate the motivation of our study. Since we make heavy use of them in this subsection, they are copied to here again:
\begin{equation}\label{QPannihilation}
\Psi^{\text{R}/\text{L}}_{e/2q, \ell}  = e^{\frac{i}{2pq}\phi^{\text{R}/\text{L}}_{j/j+1}}.
\end{equation}
This operator is non-local, as it cannot be expressed as a product of electronic operators. In the language of CFT, it is a chiral primary operator. On the other hand, $e^{i\phi^{\text{R}/\text{L}}}$ is a local operator, which acts as a simple current in the CFT description. It can be interpreted as creating/annihilating a ``trivial" quasiparticle of charge $pe$ (this is like the charge-$e$ quasiparticle in the Laughlin state, which is identified with the vacuum). Therefore, the number of chiral primaries of the edge CFT, or equivalently, the number of distinct quasiparticles living on a specific link equals to $N=2pq$. The notations in Eq. (\ref{QPannihilation}) are adopted so as to make intuitive sense when combined with the picture of the wire model, see Fig. \ref{qpoperators}: a quasiparticle can be created in two equal manner, in one case by acting $(\Psi^\text{R}_{e/2q,\ell})^\dagger$ on the $j$-th wire and creating a quasiparticle to its \textit{right}; in another case by acting $(\Psi^\text{L}_{e/2q,\ell})^\dagger$ on the $(j+1)$-th wire and creating a quasiparticle to its \textit{left}.

\begin{figure}[t!]
   \includegraphics[width=8.5cm,height=4cm ]{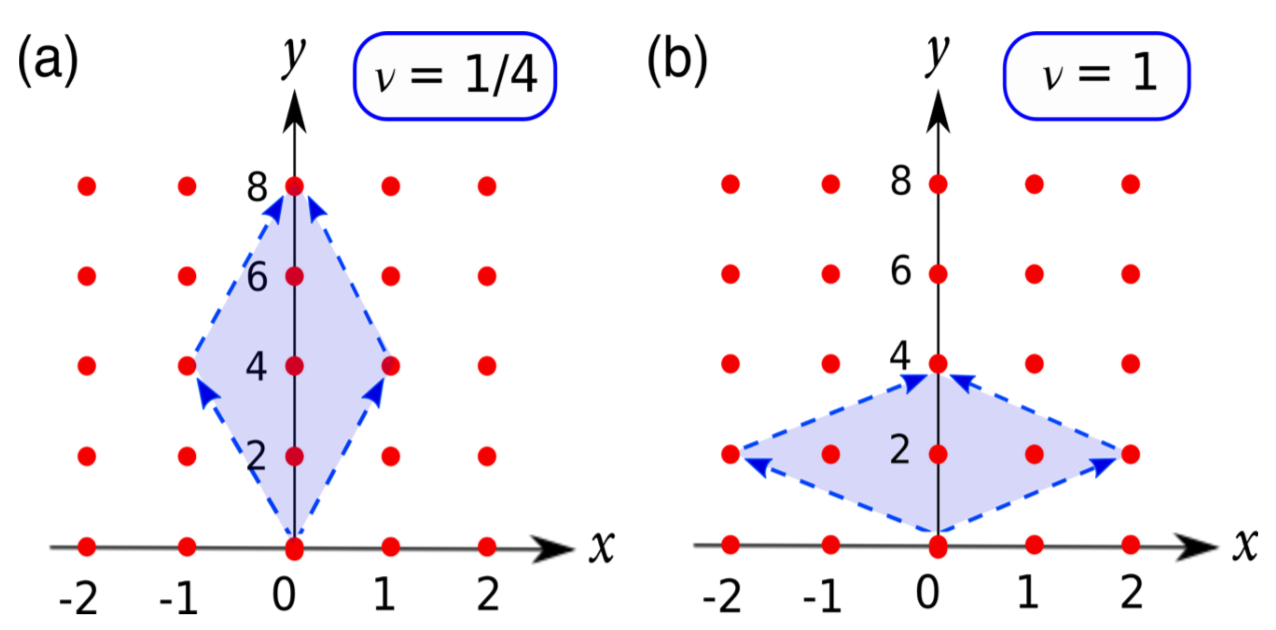}\centering
  \caption{\small{Diagrams of the allowed local scattering operators acting on a single wire for a bosonic system. Each lattice point corresponds to an operator $e^{i(x\varphi-y\theta)}$. Red dots are labeling the points $(x,y) = (r,2s)$ with $r,s\in \mathbb{Z}$, which correspond to local operators. (a) For $p=1$ and $q=2$, which gives a diagonal Laughlin state at $\nu=1/4$. (b) For $p=2$ and $q=1$, which gives a non-diagonal bosonic state at $\nu=1$. The arrows connecting the origin to the points $(x,y)=(\pm p, 2q)$ represent the creation/annihilation operators for the trivial quasiparticle, and they bind the shaded region which covers all $N=2pq$ distinct quasiparticle scattering operators. Only operators on the vertical axis are charge-neutral, while all others involve injection/removal of electrons from a single wire.}}
  \label{localoperator}
\end{figure}

Next we construct local operators that act on a single wire, and interpret their effect as scattering/moving quasiparticles from one side of the wire to another. This interpretation is also reflected in the langauge of CFT, where a physical operator is a product of a chiral and an anti-chiral vertex operators. Combination of the two would determine the partition function, as discussed in Sec. \ref{sec2.2}. Let us begin with the following operator
\begin{equation}\label{localop}
\mathcal{O}^{\{r,s\}}_j = e^{i(r\varphi_j-2s\theta_j)},
\end{equation}
which is known to be local when $r,s \in \mathbb{Z}$, as it can be constructed out of the electronic operators introduced in Eqs. (\ref{phi}) and (\ref{theta}). In this way, all local operators can be organized onto a lattice, each labeled by a point $(x,y)=(r,2s)$, as depicted in Fig. \ref{localoperator}.  Making the change of variables to chiral bosonic fields, the above local operator becomes
\begin{equation}\label{localop2}
\mathcal{O}^{\{r,s\}}_j = {\rm exp}\; \frac{i}{2pq}[(qr-ps)\phi^\text{R}_j+(qr+ps)\phi^\text{L}_j].
\end{equation}
This expression has two important implications. Firstly, on a formal ground this is connected to the partition function for the CFT of a single wire. The way that chiral/anti-chiral \textit{vertex operators} are combined to form a physical operator, as shown in Eq. (\ref{localop2}), can be translated to the way chiral/anti-chiral \textit{characters} are combined to form the modular-invariant partition function, as shown in Eq. (\ref{partition}). Specifically, by expressing $n=qr-ps$ and using the definition of $\omega$ in Eq. (\ref{Bezout}), we have $\omega n = qr+ps$ and hence the dictionary follows:
\begin{equation}
\mathcal{O}^{\{r,s\}} \leftrightarrow K_n \overline{K}_{\omega n}.
\end{equation}
Secondly, from Eq. (\ref{localop2}) one can read off the physical effect of $\mathcal{O}^{\{r,s\}}_j$, which is to scatter a quasiparticle of charge $e(qr-ps)/2q$ residing on link $j+1/2$ to another quasiparticle of charge $-e(qr+ps)/2q$ residing on link $j-1/2$. An example is depicted in Fig. \ref{qpoperators}. Notice for the special case that $r=\pm p$ and $s= \pm q$, the operator is actually creating/annihilating a ``trivial" quasiparticle of charge $pe$.  As we see next, the aforementioned implications suggest that we can learn about the scattering pattern of quasiparticles by examining the local operator $\mathcal{O}^{\{r,s\}}$, and since $\mathcal{O}^{\{r,s\}}$ is related to the partition function, the distinction between diagonal and non-diagonal CFT is also reflected in the scattering of bulk quasiparticles. 

Alongside the constraint of locality, the constraint of charge conservation also plays an important role here. It is clear from Eq. (\ref{localop}) that the local scattering operator is \textit{charged} when $r\neq 0$, since it removes $r$ electrons from the $j$-th wire. Therefore, combining locality with charge conservation, only quasiparticles with charge $pe/2q$ (or its multiples) can be scattered across a single wire, which corresponds to the operators on the vertical axis in Fig. \ref{localoperator}. For the ``diagonal" Abelian states, in which case $p=1$, all quasiparticles can be scattered across a single wire. This can also be seen from Fig. \ref{localoperator}(a), where all distinct non-trivial scattering operators lie on the vertical axis, and thus are charge-neutral. The Laughlin states, as well as the strongly-clustered states \citep{CWC17,CWC18},  belong to this category. On the other hand, for the ``non-diagonal" states with $p>1$, there exist quasiparticles, (including the minimal quasiparticle with charge $e/2q$) that cannot be scattered across just a single wire, as local operators that would scatter them require injection/removal of electrons. A representative situation is illustrated in Fig. \ref{localoperator}(b), which clearly shows the existence of charged scattering operators in the non-diagonal case. 

Having said that, by no means do we imply that quasiparticles with charge other than (multiples of) $pe/2q$ cannot move at all in the bulk of non-diagonal states. Though they cannot hop across just a \textit{single} wire, they can indeed hop across \textit{two}. For instance, the following local operator:
\begin{equation}\label{scattertwowire}
\mathcal{O}^{\{-r,s\}}_{j-1}\mathcal{O}^{\{r,s\}}_j \propto {\rm exp}\;i[\frac{(qr-ps)}{2pq}(\phi^\text{R}_j-\phi^\text{L}_{j-1})]
\end{equation}
would scatter a quasiparticle with charge $e(qr-ps)/2q$ from link $j+1/2$ to link $j-3/2$. Note that in writing Eq. (\ref{scattertwowire}) we have used the fact that $\Theta_{j-1/2}=\phi^\text{R}_{j-1}-\phi^\text{L}_{j}$ has been pinned at $2\pi n$ ($n\in \mathbb{Z}$) as the bulk has been gapped, so there is a numerical constant that can be factored out, which explains the proportionality sign. This operator is charge-neutral, because while $r$ electrons are removed from the $j$-th wire, $r$ electrons are injected into the $(j-1)$-th wire, so the the minimal quasiparticle can indeed move across two wires given an inter-wire tunneling interaction, which is local and charge-preserving. 

To summarize, non-diagonal quantum Hall states can be distinguished from the diagonal states in terms of the scattering pattern of bulk quasiparticle in the wire model. While all quasiparticles in the diagonal states can hop across a single wire, certain quasiparticles (including the minimal quasiparticle) in the non-diagonal states are only allowed to hop across two wires at a time. This then differentiates two types of quasiparticles: ones which live on the \textit{even} links, and the others which live on the \textit{odd} links. As we will show in Sec. \ref{sec3}, they can be associated to the $\mathbf{e}$-type and $\mathbf{m}$-type anyons in the $\mathbb{Z}_p$ toric code (also known as the $\mathcal{D}(\mathbb{Z}_p)$ quantum double model), thus allowing us to assign a $\mathcal{D}(\mathbb{Z}_p)$ neutral sector to the non-diagonal states. After establishing this relation, we will also rigorously address the difference between non-diagonal states and strongly-clustered states, from the perspectives of intrinsic topological order and symmetry-enriched topological order. 

\subsection{Fermionic states}\label{sec2.4}

In the above discussion we have been focusing on bosonic non-diagonal states. Here we explain that a similar constrained pattern of quasiparticle motion also arise in fermionic QH states. Moreover, in an analogous setting of 2d weak topological superconductor, vortices have a similar constrained motion that can be understood in terms of fractional Josephson effect. 

\subsubsection{Non-diagonal quantum Hall states}\label{sec2.4.1}
The coupled wire construction for the fermionic non-diagonal state is detailed in Appendix \ref{secappendixa}. The inter-wire coupling is essentially the same, but due to the non-local nature of fermion, which requires attaching a Jordan-Wigner string to the bosonized electron operator, the filling fraction is modified to
\begin{equation}
\nu = \frac{p}{p+2q}.
\end{equation}
In fact, most changes from the bosonic case to the fermionic case can be accounted for by substituting $2q \mapsto p+2q$. The annihilation operator for the minimal quasiparticle on link $\ell=j+1/2$ is
\begin{equation}
\Psi^{\text{R}/\text{L}}_{e/(p+2q), \ell} = e^{\frac{i}{p(p+2q)}\phi^{\text{R}/\text{L}}_{j/j+1}},
\end{equation}
where the quasiparticle carries charge $e/(p+2q)$. Here $\phi^{\text{R}/\text{L}}$ is the chiral bosonic field of a circle CFT with compactification radius $R=\sqrt{p/(p+2q)}$, and $e^{\pm i\phi^{\text{R}/\text{L}}}$ creates/annihilates a charge-$pe$ trivial quasiparticle. A physical operator that scatters a quasiparticle across a single wire takes the following form:
\begin{equation}\label{fermioniclocalop2maintext}
\mathcal{O}^{\{r,s\}}_j = \text{exp} \frac{i}{p(p+2q)}[(qr-ps)\phi^\text{R}_j+(qr+ps+pr)\phi^\text{L}_j], 
\end{equation}
with $r,s\in \mathbb{Z}$. To avoid confusion, let us be more clear about our terminology: an operator is physical (and thus allowed) in the sense that it can be expressed in terms of electronic operators. For bosonic states, a physical operator is equivalently a local operator, as bosons are local objects. Thus, we have used the terms ``physical" and "local" interchangeably in the earlier discussion. However, since fermions are non-local, a distinction should be made here. 

\begin{figure}[t!]
   \includegraphics[width=8.5cm,height=4cm ]{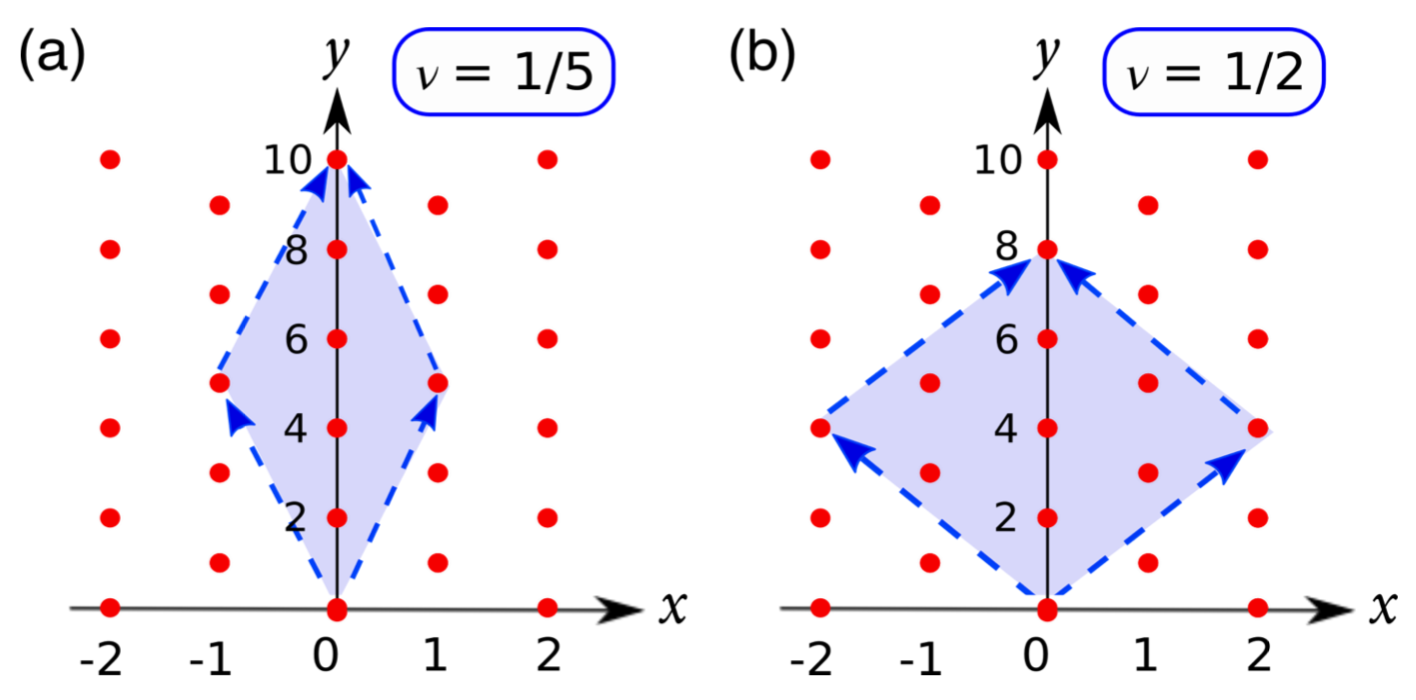}\centering
  \caption{\small{Diagrams of the allowed physical scattering operators acting on a single wire for a fermionic system. Each lattice point corresponds to an operator $e^{i(x\varphi-y\theta)}$. Red dots are labeling the points $(x,y) = (r,r+2s)$ with $r,s\in \mathbb{Z}$, which correspond to physical operators. (a) For $p=1$ and $q=2$, which gives rise to the diagonal Laughlin state at $\nu=1/5$. (b) For $p=2$ and $q=1$, which gives rise to a non-diagonal fermionic state at $\nu=1/2$. The arrows connecting the origin to the points $(x,y)=(\pm p, p+2q)$ represent the creation/annihilation operators for the trivial quasiparticle, and they bind the shaded region which covers all $N=p(p+2q)$ distinct quasiparticle scattering operators. Only operators on the vertical axis are charge-neutral, while all others involve injection/removal of electrons from a single wire. In particular, operators with odd $x$ (or $r$) also change the fermion-parity.}}
  \label{flocaloperator}
\end{figure}

Physical operators in the fermionic state can be organized into a lattice as depicted in Fig. \ref{flocaloperator}, which is analogous to Fig. \ref{localoperator} for the bosonic state. Notice that the lattice here is in a checker-board pattern because a physical operator is now attached to a Jordan-Wigner string. The action of $\mathcal{O}^{\{r,s\}}_j$ is to scatter a quasiparticle of charge $e(qr-ps)/(p+2q)$ across the $j$-th wire to become a quasiparticle of charge $-e(qr+ps+pr)/(p+2q)$. Analogous to the bosonic case, this operator violates charge conservation when $r\neq 0$, so the associated scattering process is forbidden in the presence of $U(1)$ charge symmetry. For the fermionic states with $p>1$, the minimal quasiparticle clearly cannot hop across just a single wire. Its motion is constrained to hop across two wires at a time, which can be achieved by exchanging electrons between the two wires. Again, quasiparticles on the even links shall be distinguished from those on the odd links, which is considered to be the defining feature of a non-diagonal quantum Hall state. As in the bosonic case, this would allow us to associate the quasiparticles to anyons in the $\mathbb{Z}_p$ toric code.

We care to describe the fermionic case not only because it is physically more relevant, but also because there is a subtle difference between it and the bosonic case. For the bosonic non-diagonal states, we have emphasized the importance of charge conservation in constraining the motion of quasiparticles. However, for a fermionic system, one can also talk about the conservation of \textit{fermion parity}, which could play an additional role. Due to the non-locality of fermionic electrons, $\mathbb{Z}_2$ fermion-parity symmetry is more robust than the $U(1)$ charge symmetry. For fermionic states, the physical scattering operator $\mathcal{O}^{\{r,s\}}_j $ with $r\in 2\mathbb{Z}+1$ violates not only charge conservation but also fermion-parity conservation. Therefore, the constrained motion of \textit{some} quasiparticles in the fermionic state is more robustly protected by the fermion-parity symmetry. Having said that, in general the fermion-parity symmetry cannot completely replace the role of charge symmetry. We will elaborate more on this issue when we discuss the symmetry-enrichment of bosonic and fermionic non-diagonal states in Sec. \ref{sec3}. Before moving on, it is instructive to take a digression and consider a setting different from the FQH state, where fermion-parity symmetry \textit{alone} can constrain the motion of low-energy excitations in the way we have just discussed. 

\subsubsection{Fractional Josephson effect}\label{sec2.4.2}
Let us consider a wire model consisting of one-dimensional superconductors, each described by a single-channel quantum wire with attractive interaction. With superconductors, charge is no longer conserved while fermion-parity still is. Instead of coupling wires to form a quantum Hall state, a two-dimensional superconductor is formed by locking the pairing phases between neighboring wires. As discussed in Refs. \citep{CWC17, CWC18}, such a wire has two distinct phases: one being the ``strongly-paired" phase where effectively all electrons are bound to form Cooper pairs, in which case the wire is a 1d \textit{trivial} superconductor. Another phase is the ``weakly-paired" phase described by the coexistence of unpaired electrons and Cooper pairs, in which case the wire is a 1d \textit{topological} superconductor, and has been shown to adiabatically connect to the Luttinger liquid phase. Here we are concerned with how superconducting vortices, which are analogs of the quasiparticles in the quantum Hall setting, can move around in the coupled wire model when the constituting wires are either trivial or topological superconductors. The minimal vortex carries flux $h/2e$, around which the pairing phase $\Theta_{\rm sc}$ is advanced by $2\pi$. When the wires are trivial, or in the ``strongly-paired" phase, the vortex has no issue tunneling across a single wire, because the wire contains only charge-$2e$ Cooper pairs which are local with respect to the vortex. As the vortex tunnels across a trivial superconductor, and induces a $2\pi$ phase slip, the wire simply returns back to its original state due to the ordinary Josephson effect. 

\begin{figure}[b!]
   \includegraphics[width=8cm,height=4.1cm ]{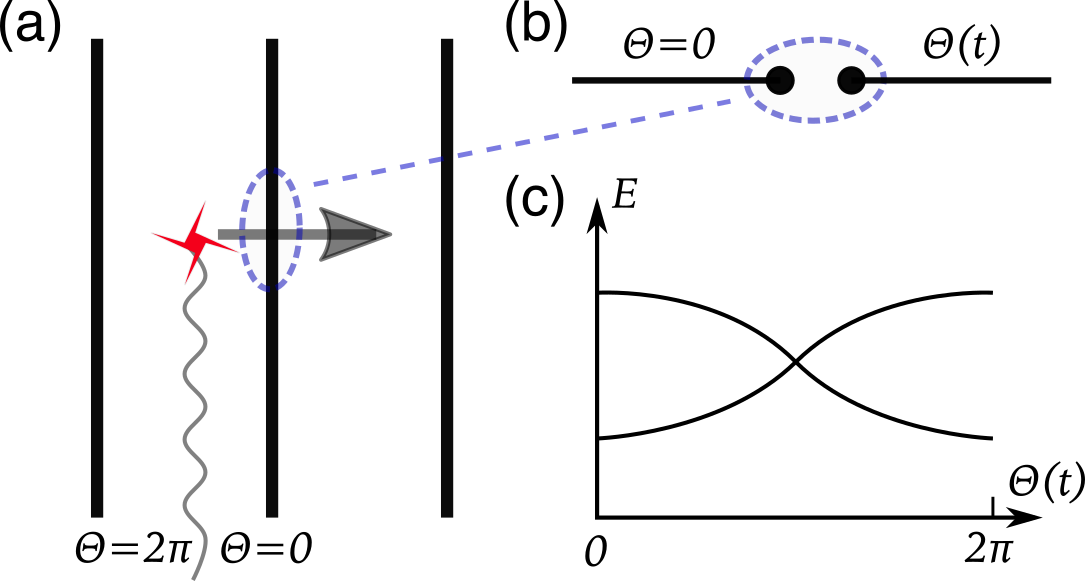}\centering
  \caption{\small{Fractional Josephson effect in weak topological superconductor. (a) shows a wire model for the weak topological superconductor, with individual wires being 1d topological superconductors. Between two wires live a single vortex excitation, around which the pairing phase $\Theta_{\rm sc}$ is advanced by $2\pi$. When the vortex hops across the middle wire, its associated branch cut is also dragged across the wire to induce a $2\pi$ phase slip there. (b) illustrates the above process by modeling the place where the vortex crosses the wire as a Josephson junction. There the topological superconductor is cut open and hosts two Majorana modes that define a fermion parity in the circled region. (c) shows the evolution of energy levels of the coupled Majorana modes as a function of phase difference. A $2\pi$ phase slip leads to a change of fermion parity in the ground state.}}
  \label{josephson}
\end{figure}

Things are different when the wires are 1d topological superconductors, which when coupled together form the \textit{weak topological superconductor}. A possible material realization of this setup has been proposed for a thin slab of Sr$_2$RuO$_4$ with enhanced pairing instability for the quasi-1D band \cite{Raghu2010weakTSC, Hughes2014weakTSC}. In this case there are unpaired electrons in each wire, which are non-local with respect to the $h/2e$-vortex. Consequently, tunneling the vortex across the wire would lead to the fractional Josephson effect as illustrated in Fig. \ref{josephson}. The tunneling process can be modeled by cutting the wire open at the place where the process happens, and since the wire is topological, each open end hosts a Majorana mode. The Majorana modes are coupled in the Josephson junction and together defines a fermion parity for the weak link. As predicted by Kitaev \citep{KitaevChain}, a $2\pi$ phase slip leads to a swtich of fermion parity for the ground state, and thus injection/removal of an electron is required in order to move a single vortex across the wire. The fermion parity symmetry for a single wire thus forbids the minimal vortex from moving across it.

In this situation, there are two ways for vortices to move in the bulk. One way is for a \textit{double-vortex} to tunnel across a single wire, which leads to a total $4\pi$ phase slip that restores the fermion parity of the wire. Also, from the perspective of locality the $h/e$-vortex is local with respect to everything in the topological superconductor, and hence should be allowed to tunnel across. Alternatively, a single $h/2e$-vortex can tunnel across \textit{two} wires at a time, as this can be achieved by exchanging fermion parity between the two wires. These features are analogous to what we have advertised for the non-diagonal QH states, and later we will see that these give rise to an interpretation of the $h/2e$ vortices as anyons in the toric code. We will further comment on this similarity, as well as an important difference in this regard, when compared with the non-diagonal state in Sec. \ref{sec3.2}. 

Aside from the scattering pattern of low-energy excitations, there is yet another revealing similarity with the original wire model for QH states that is worth mentioning. Just as the 1d topological superconductors that host Majorana end modes, the quantum wires used for constructing non-diagonal QH states actually host $\mathbb{Z}_p$ parafermion end modes. This is related to the inter-wire coupling discussed in Sec. \ref{sec2.1}, which preserves the particle number mod $p$ of each wire. Coupling together these modes that appear at the top/bottom edge would lead to an edge theory that is fundamentally different from the one describing the left/right side edge. We analyze this in detail in Sec. \ref{sec4}. Given discrete translation symmetry in the bulk, it leads to a gapless (for $p=2,3$ at least) theory for the top/bottom edge that is even richer than the side edge already addressed in Sec. \ref{sec2.2}. Next, let us unveil more interesting physics from the bulk perspective first, using the tools we have just developed, which would eventually guide us to a complete description for the edge of non-diagonal states. 

\section{\label{sec3}Neutral sector as $\mathbb{Z}_p$ toric code}
In this section we first calculate the braiding statistics of quasiparticles in non-diagonal quantum Hall states, so as to reveal a ``hidden" $\mathbb{Z}_p$ topological order that can be attributed to the neutral sector. It will be explained later that this additional topological order originates from \textit{symmetry-enrichment} \cite{Ran2013SET, Hung2013SET, Lu2016SET, Cheng2016translationSET, Chen2017SET}, which ultimately distinguishes the non-diagonal states from the strongly-clustered states. For simplicity in exposition, the following discussion would mostly refer to the bosonic states. Further specification would be made when the fermionic case is worth a distinction.

As demonstrated in the wire model for the $\nu=p/2q$ non-diagonal state, quasipaticles are created/annihilated by the vertex operators in Eqs. (\ref{QPcreation}) and (\ref{QPannihilation}), so the fusion algebra is simply Abelian. To characterize the topological order, we focus our attention on the braiding statistics. At first sight, it appears like the topological data resembles to those of the Laughlin states. Indeed, the non-trivial Abelian quasiparticle with minimal charge $e/2q$ also exists in the Laughlin state of charge-$pe$ bosons at filling $\nu=1/2pq$ (also known as the strongly-clustered state). Nevertheless, as we have noted before, quasiparticles in the non-diagonal states have constrained motion in the bulk, which differentiates excitations on the even links from those on the odd links. The result of braiding depends on whether two quasiparticles live on links of the the same type or not, and as we will see in Sec. \ref{sec3.1}, the result can be understood in the context of the $\mathcal{D}(\mathbb{Z}_p)$ quantum double model by associating quasiparticles on even/odd links to $\mathbf{e}/\mathbf{m}$-particles respectively. The $\mathcal{D}(\mathbb{Z}_p)$ quantum double model has a $\mathbb{Z}_p$ topological order, and is also known as the $\mathbb{Z}_p$-generalization of Kitaev's toric code (which has $p=2$) \cite{Teo2015twistliquid, Teo2016AS, Kitaev2006exactly, Kitaev2003TQC, Barkeshli2019AS}. 

It is important to notice that the distinction between even links and odd links originates from $U(1)$ charge symmetry. Specific examples of non-diagonal states are analyzed in Sec. \ref{sec3.2} to demonstrate how this would affect the quasiparticle spectrum. Besides, the $\mathbf{e}\leftrightarrow\mathbf{m}$ anyonic relabeling is related to the $\mathbb{Z}$ translation in the coupled wire model, which lead us to eventually identify the non-diagonal states, with a $\mathbb{Z}_p$ toric code in the neutral sector, as a $U(1)\times \mathbb{Z}$ symmetry-enriched topological (SET) order. Furthermore, the ``gauging" of anyonic symmetry can be physically realized in the wire model as the proliferation of dislocation defects, which are sudden terminations of wires in the bulk. We explain these in Sec. \ref{sec3.3}. 

\subsection{Braiding statistics}\label{sec3.1}

The braiding statistics is encoded in the quasiparticle operators studied in Sec. \ref{sec2.3}. Let us begin with a quasiparticle of charge $n e/2q$ on link $\ell=j+1/2$. Here we denote $n=qr-ps$ for some $r, s \in \mathbb{Z}$. Following Eq. (\ref{scattertwowire}), the local operator that transfers this quasiparticle from link $\ell$ to link $\ell-2\mathcal{N}$ can be written as 
\begin{equation}
\begin{split}
&\prod_{\eta=0}^{\mathcal{N}-1}\mathcal{O}^{\{-r,s\}}_{j-1-2\eta}(x) \mathcal{O}^{\{r,s\}}_{j-2\eta}(x)\\
&= e^{i\frac{n}{2pq}[\phi^\text{R}_j (x) - \phi^\text{L}_{j-2\mathcal{N}+1}(x)]}\;\;\times\\
&\;\;\;\;\;\prod_{\mu=1}^{2\mathcal{N}-1} e^{i\frac{n}{2pq}\Theta_{\ell-\mu}(x)} \times \prod_{\eta = 0}^{\mathcal{N}-1}e^{-i\frac{r_0 n}{p}\Theta_{\ell-2\eta-1}(x)}
\end{split}
\end{equation} 
with $r_0$ in the last term defined in Eq. (\ref{Bezout}). The first term with the chiral fields clearly generates the anticipated scattering process. Here we have made explicit the wire coordinate $x$, and retain the link variables $\Theta(x)$ which are essential to deducing the braiding phase. Notice that in the last equality the second term is contributed by every link between the first ($\ell$) and the last ($\ell-2\mathcal{N}$), while the third term is contributed only by the links whose indices are of the same parity as $\ell$. 

\begin{figure}[t!]
   \includegraphics[width=7.5cm,height=5.3cm ]{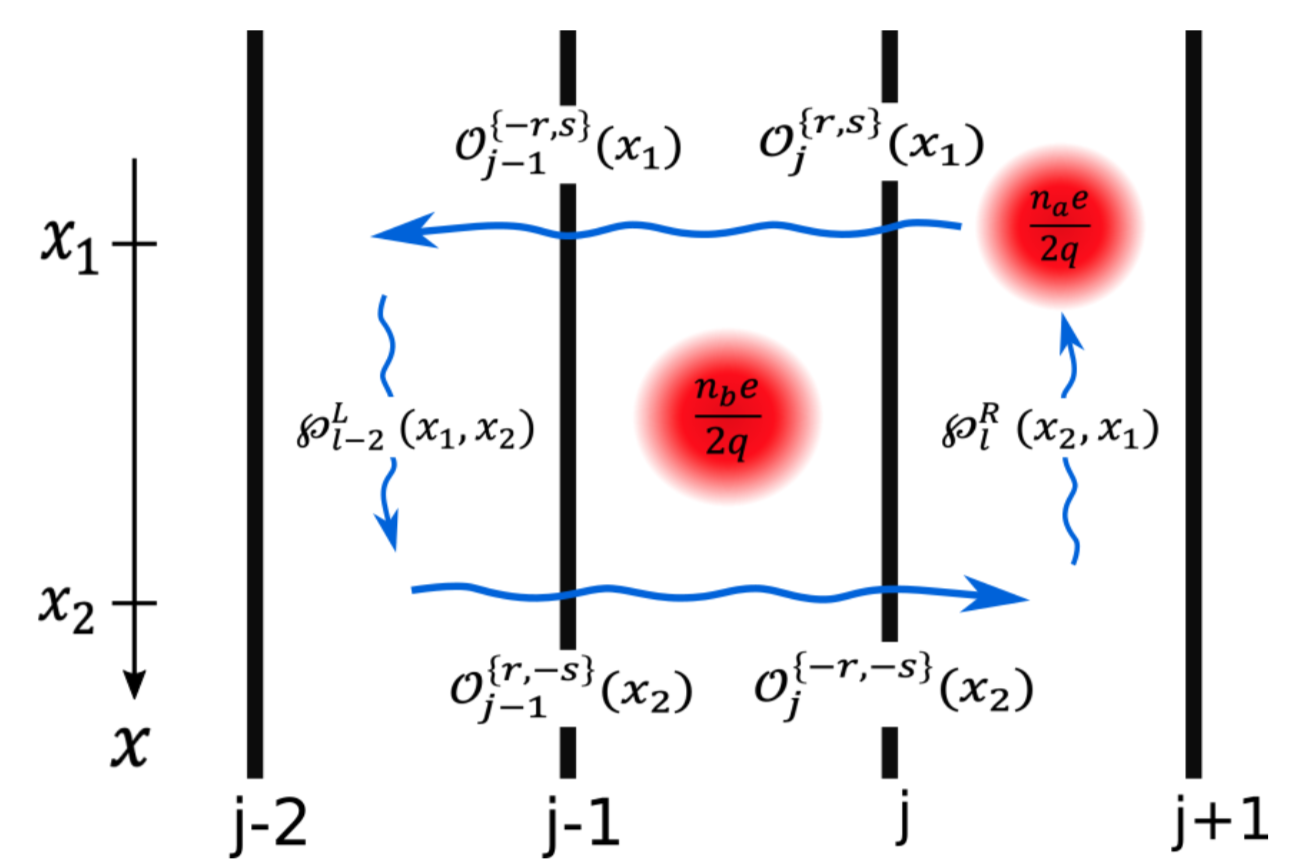}\centering
  \caption{\small{Schematic illustration of a braiding process between two quasiparticles, one living on the even links and another on the odd links.}}
  \label{braiding}
\end{figure}

In order to consider a closed loop for a braiding process one also needs to move a quasiparticle on link $\ell$ \textit{along} the wire direction, say from $x_1$ to $x_2$. This is accomplished by the following operator:
\begin{equation}
\begin{split}
\wp^{\text{R}/\text{L}}_{\ell}(x_1,x_2) &= {\rm exp}\;i\frac{n}{2pq}[\phi^{\text{R}/\text{L}}_{j/j+1}(x_1)-\phi^{\text{R}/\text{L}}_{j/j+1}(x_2)]\\
& = {\rm exp}\;i\frac{n}{2pq}\int_{x_2}^{x_1}dx \;  \partial_x \phi^{\text{R}/\text{L}}_{j/j+1}(x)\;. 
\end{split}
\end{equation}
This is indeed a local operator, as the last equality suggests that it can be expressed in terms of bare electron densities and currents. With these established, we can consider transferring a quasiparticle around a closed loop by local operators. To be specific, let us set up a coordinate $(\ell, x)$ for a quasiparticle at position $x$ on link $\ell$, then we want to consider the loop  $\mathcal{C}: (\ell, x_1) \rightarrow (\ell-2\mathcal{N}, x_1) \rightarrow (\ell-2\mathcal{N}, x_2) \rightarrow (\ell, x_2) \rightarrow (\ell, x_1)$. An example is depicted in Fig. \ref{braiding}. 

It is now clear that the following phase is picked up after completing the loop $\mathcal{C}$:
\begin{equation}
\begin{split}
&\prod_{\mu=1}^{2\mathcal{N}-1} e^{i\frac{n}{2pq}[\Theta_{\ell-\mu}(x_1)-\Theta_{\ell-\mu}(x_2)]}\\ & \times \prod_{\eta = 0}^{\mathcal{N}-1}e^{-i\frac{r_0 n}{p}[\Theta_{\ell-2\eta-1}(x_1)-\Theta_{\ell-2\eta-1}(x_2)]}. 
\end{split}
\end{equation}
Recall that the bulk is gapped so that $\Theta$'s are pinned at integer multiples of $2\pi$, while quasiparticle excitations from the ground state correspond to $2\pi$-kinks. Thus the first term is contributed by every enclosed quasiparticle, while the second term is contributed only by the enclosed quasiparticles that live on the links with parity \textit{different} from that of $\ell$. Hence the braiding phase for two quasiparticles $\mathbf{a}$ and $\mathbf{b}$, with charge $n_a e/2q$ and $n_b e/2q$ respectively, is encoded in the following matrix:
\begin{equation}\label{braidingmatrix}
M_{\bar{a}\bar{b}} = e^{2\pi i\frac{n_a n_b }{2pq}} \begin{pmatrix}
1 & e^{-2\pi i \frac{r_0 n_a n_b}{p}} \\ e^{-2\pi i \frac{r_0 n_a n_b}{p}} & 1
\end{pmatrix}, 
\end{equation}
with the matrix index $\bar{a}=1/2$ for the quasiparticle $\mathbf{a}$ living on the odd/even links. The braiding statistics for the fermionic state is obtained by substituting $2q$ with $p+2q$ in the above discussion (in both cases $r_0$ is defined by $qr_0-ps_0=1$). The first factor gives the mutual statistics between quasiparticles of the same type, namely for those living on links of the same parity. The same braiding statistics describes a strongly-clustered state, which is essentially a Laughlin state of charge-$pe$ particles at filling $\nu_{pe}=1/2pq$ (or $\nu_{pe}=1/p(p+2q)$ in the fermionic case). For $p=1$, this is the full story because even and odd links need not be distinguished. However, for $p>1$, which corresponds to a non-diagonal state, the topological order is richer. The second factor in Eq. (\ref{braidingmatrix}) is not an identity matrix for $p>1$, and since $r_0$ is by definition coprime to $p$, it is actually the braiding matrix for the $\mathcal{D}(\mathbb{Z}_p)$ quantum double model. Below, we briefly overview this well-known topological order. 

\subsubsection{The $\mathcal{D}(\mathbb{Z}_p)$ quantum double model}
The $\mathcal{D}(\mathbb{Z}_p)$ quantum double model is a non-chiral Abelian topological order that can be realized in the deconfined phase of the $\mathbb{Z}_p$ discrete gauge theory in 2+1D. Alternatively, it can be characterized by a two-component Chern-Simons theory with the following Lagrangian \cite{WenZee1992CS}: 
\begin{equation}
\mathcal{L} = \frac{\epsilon^{\mu\nu\lambda}}{4\pi}\vec{\alpha}^T_\mu K \partial_\nu \vec{\alpha}_{\lambda}+ \vec{\alpha}^T_\mu \vec{j}_\mu, 
\end{equation}
where $\vec{\alpha}^T = (\alpha^1, \alpha^2)$ represents the internal $U(1)^2$ gauge field and $\vec{j}$ is the quasiparticle current. The $K$-matrix which encodes all the topological data is
\begin{equation}
K = p\sigma_x = \begin{pmatrix}
\;\;0\; & \;\;p\;\; \\ \;\;p\;& \;\;0\;\; 
\end{pmatrix}.
\end{equation}
The chiral central charge for this phase is $c\propto \text{tr}(K) = 0 $. A quasiparticle is labeled by a two-component vector $\vec{t}$ defined on the so-called anyon integral lattice $\Gamma^* = \mathbb{Z}^2$, while the sublattice $\Gamma = K\Gamma^*$ consists of the states that are local particles which braid trivially with all quasiparticles, and thus belong to the identity topological sector. Hence, distinct quasiparticles are defined on the quotient lattice $\Gamma^*/\Gamma$, which in this case has $p^2$ points. There are two types of minimal quasiparticle, which is the $\mathbf{e}$-particle with $\vec{t}^T = (1,0)$ and the $\mathbf{m}$-particle with $\vec{t}^T = (0,1)$. The $p^2$ distinct quasiparticles in the $\mathcal{D}(\mathbb{Z}_p)$ quantum double model can thus be labeled by $\mathbf{e}^\alpha\mathbf{m}^\beta$, where $0\leq \alpha < p$ and $0\leq \beta < p$.  

The complete topological information is specified by the fusion algebra and the braiding statistics. The fusion algebra is Abelian: 
\begin{equation}\label{TCfusion}
\mathbf{e}^{\alpha_1}\mathbf{m}^{\beta_1} \times \mathbf{e}^{\alpha_2}\mathbf{m}^{\beta_2} = \mathbf{e}^{\alpha_1+\alpha_2}\mathbf{m}^{\beta_1+\beta_2}. 
\end{equation}
Notice that $\mathbf{e}^{p}=\mathbf{m}^{p}=\mathds{1}$ is the trivial quasiparticle. The self and mutual statistics are encoded in the $\mathcal{T}$ and $\mathcal{S}$ matrices (the fusion algebra also follows from $\mathcal{S}$ through the Verlinde formula), and in the $K$-matrix formulation they are given by:
\begin{equation}\label{STmatrices}
\mathcal{T}_{\mathbf{a}\mathbf{b}}= \delta_{\mathbf{a}\mathbf{b}}e^{\pi i\vec{a}^TK^{-1}\vec{a}}, \;\; \mathscr{D}\mathcal{S}_{\mathbf{a}\mathbf{b}} = e^{2\pi i\vec{a}^TK^{-1}\vec{b}}. 
\end{equation}
Here $\vec{a}, \vec{b} \in \Gamma^*/\Gamma$ are the vectors in the anyon lattice labeling the two quasiparticles $\mathbf{a}$ and $\mathbf{b}$, and $\mathscr{D}=\sqrt{\abs{{\rm det}K}}=p$ is the total quantum dimension. It follows that $\mathbf{e}$-particles and $\mathbf{m}$-particles are all self-bosons (they have trivial self-exchange statistics), while $\mathbf{e}^{\alpha}$ and $\mathbf{m}^\beta$ have a non-trivial braiding phase of $e^{2\pi i\frac{\alpha\beta}{p}}$. The case with $p=2$ has four anyons: $\mathds{1}, \mathbf{e}, \mathbf{m}$ and $\psi=\mathbf{em}$ (a composite fermion), which is exactly the topological order in Kitaev's toric code \cite{Kitaev2003TQC, Kitaev2006exactly}. 

There is an important global symmetry in the $\mathcal{D}(\mathbb{Z}_p)$ quantum double model, known as the $\mathbb{Z}_2$ $\mathbf{e}$-$\mathbf{m}$ anyonic symmetry. More precisely, it is an anyon \textit{relabeling} symmetry that interchanges the $\mathbf{e}$-particles with the $\mathbf{m}$-particles, leaving the fusion rules and braiding statistics invariant. In the $K$-matrix formalism, the $\mathbb{Z}_2$ anyonic symmetry is implemented by acting $\sigma_x$ on the anyon lattice $\Gamma^*$, or equivalently by transforming $K \rightarrow \sigma_x^T K \sigma_x =K$. As the $K$-matrix is left invariant, it is clear that all topological information is left invariant. Anyonic symmetry is of physical importance because a non-Abelian phase can be obtained from gauging the anyonic symmetry in an Abelian phase \cite{Teo2015twistliquid, Teo2016AS, Barkeshli2019AS}. 

\subsubsection{$\mathcal{D}(\mathbb{Z}_p)$ in non-diagonal QH states}
Getting back to our original discussion, one realizes that the braiding statistics in Eq. (\ref{braidingmatrix}) can be understood by viewing the (bosonic) non-diagonal quantum Hall state as consisting of a $U(1)_{2pq}$ charge sector and a $\mathcal{D}(\mathbb{Z}_p)$ neutral sector. The net braiding phase is obtained by adding the phase in the charge sector and the phase in the neutral sector. A quasiparticle of charge $n_a e/2q$ excited on an even link can be labeled by $(n_a, \mathbf{e}^{-r_0 n_a})$, while a quasiparticle of charge $n_b e/2q$ excited on an odd link can be labeled by $(n_b, \mathbf{m}^{n_b})$. A generic quasiparticle, which can be viewed as a composite of quasiparticles from even and odd links, is then denoted as
\begin{equation}\label{quasiparticlelabel}
(n_a+n_b\;,\; \mathbf{e}^{-r_0 n_a}\mathbf{m}^{n_b}). 
\end{equation}
The first component represents the electric charge (in unit of $e/2q$), while the second component represents the neutral sector and obeys $\mathbf{e}^{p}=\mathbf{m}^{p}=\mathds{1}$. The interpretation of a $\mathbb{Z}_p$ toric code in the neutral sector also implies that the $\mathbf{e}$-$\mathbf{m}$ anyonic symmetry is concretely realized in the wire model as the discrete translation symmetry by one wire. More precisely, here the $\mathbb{Z}_2$ anyonic symmetry interchanges $\mathbf{e}^{-r_0} \leftrightarrow \mathbf{m}$, with $r_0$ defined in Eq. (\ref{Bezout}). 


Next, let us analyze some specific examples of non-diagonal states, which would familiarize ourselves with the connection to the $\mathbb{Z}_p$ toric code just advertised. Moreover, they highlight the importance of symmetry considerations, particularly the $U(1)$ charge symmetry, for characterizing the topological order of non-diagonal states. There are also exceptional cases in which the fermion-parity symmetry can replace the role of charge symmetry.

\subsection{Examples}\label{sec3.2}
\subsubsection{Bosonic state}
We first study a representative example of bosonic non-diagonal states, with $p=2$ and $q=1$, which occur at filling $\nu=1$. According to the discussion in Sec. \ref{sec2.3}, each link hosts four distinct quasiparticle excitations, which have charge $0$, $e/2$, $e$ and $3e/2$ respectively. Figure \ref{localoperator}(b) summarizes the possible scattering operators. In a system with charge conservation, only the charge-$e$ quasiparticle (and the trivial quasiparticle) can hop across a single wire, while the charge-$e/2$ and charge-$3e/2$ excitations cannot, so those on the even links are regarded as different from those on the odd links. A quasiparticle excitation composed of a charge-$e/2$ excitation on the even link and a charge-$e/2$ excitation on the odd link, hence with total charge $e$, is then distinct from a single charge-$e$ excitation on either the even or odd link. Using the presentation introduced in Eq. (\ref{quasiparticlelabel}), we have the following quasiparticle spectrum:
\begin{equation}\label{spectrum}
\begin{alignedat}{4}
&\text{charge }0&&:\;\;\;(0,\mathds{1}),\;&&&(0,\mathbf{em})\\
&\text{charge }e/2&&:\;\;\;(1, \mathbf{e}),&&&(1, \mathbf{m})\\
&\text{charge }e&&:\;\;\;(2, \mathds{1}),\;&&&(2, \mathbf{em})\\
&\text{charge }3e/2&&:\;\;\;(3, \mathbf{e}),&&&(3, \mathbf{m})\\
\end{alignedat}
\end{equation}
The first component labels the $U(1)_4$ charge sector, and the second component labels the $\mathcal{D}(\mathbb{Z}_2)$ neutral sector. It is important to notice that, from the point of view of \textit{intrinsic} topological order, $(2,\mathbf{em})$ should really be treated as a trivial quasiparticle due to the trivial self- and mutual-statistics, which would in turn reduce the spectrum down to only four distinct quasiparticles. Specifically, by fusing with $(2,\mathbf{em})$, $(1, \mathbf{m})$ would be identified with $(3,\mathbf{e})$, $(3, \mathbf{m})$ would be identified with $(1,\mathbf{e})$, and $(0, \mathbf{em})$ would be identified with $(2,\mathds{1})$. From this perspective, it seems unnecessary to assign a neutral sector, as $\mathbf{m}$-particles can be identified with $\mathbf{e}$-particles. The intrinsic topological order in this example is thus the same as the strongly-paired state, which only has the $U(1)_4$ charge sector. 

However, the importance of the neutral sector becomes clear from the \textit{symmetry-enriched} perspective. In particular, the constrained motion of quasiparticles in the wire model is tied up with the $U(1)$ charge symmetry, which motivates us to study the non-diagonal states in the presence of charge conservation. This then requires us to distinguish quasiparticles with different electric charge, and forbid us from identifying $\mathbf{e}$-particles with $\mathbf{m}$-particles as above. A similar discussion applies to a generic bosonic non-diagonal state.

\subsubsection{Fermionic state}
Next we study a special example of fermionic non-diagonal states, with $p=2$ and $q=1$, which occur at filling $\nu=1/2$. This state is presumably more relevant experimentally, and moreover, it is exceptional from the symmetry perspective. Unlike bosonic states, the $\mathcal{D}(\mathbb{Z}_2)$ neutral sector needs not be protected by the $U(1)$ charge symmetry. Instead, the $\mathbb{Z}_2$ fermion-parity symmetry suffices to distinguish $\mathbf{e}$-particles on even links from the $\mathbf{m}$-particles on odd links.  To see this, we list out the quasiparticle spectrum:
\begin{equation}\label{fermionspectrum}
\begin{alignedat}{4}
&\text{charge }0&&:\;\;\;(0,\mathds{1}),\;&&&(0,\mathbf{em})\\
&\text{charge }e/4&&:\;\;\;(1, \mathbf{e}),&&&(1, \mathbf{m})\\
&\text{charge }e/2&&:\;\;\;(2, \mathds{1}),\;&&&(2, \mathbf{em})\\
&\text{charge }3e/4&&:\;\;\;(3, \mathbf{e}),&&&(3, \mathbf{m})\\
&\text{charge }e&&:\;\;\;(4,\mathds{1}),\;&&&(4,\mathbf{em})\\
&\text{charge }5e/4&&:\;\;\;(5, \mathbf{e}),&&&(5, \mathbf{m})\\
&\text{charge }3e/2&&:\;\;\;(6, \mathds{1}),\;&&&(6, \mathbf{em})\\
&\text{charge }7e/4&&:\;\;\;(7, \mathbf{e}),&&&(7, \mathbf{m})
\end{alignedat}
\end{equation}
The first component labels the $U(1)_8$ charge sector, and the second component labels the $\mathcal{D}(\mathbb{Z}_2)$ neutral sector. These quasiparticles can move in the bulk by the physical operators summarized in Fig. \ref{flocaloperator}(b). Using the fermionic version of Eq. (\ref{braidingmatrix}), one can check that $(4,\mathbf{em})$ braids trivially with all quasiparticles. However, strictly speaking it does not belong to the identity sector as it carries topological spin $-1$. In fact, $(4,\mathbf{em})$ corresponds to the physical electron. For a fermionic topological order, the physical electron is usually included in the counting of topological excitations (this is known as fermion-parity grading). Thus the above spectrum is complete and irreducible. To put it another way, the fermion-parity symmetry ensures the distinction between $\mathbf{e}$-particles and $\mathbf{m}$-particles, as turning one into another would require a switch in fermion-parity. 

It is easy to verify that the distinction between even and odd links is robust for all $p=2$ non-diagonal states (\textit{i.e.} $q$ can be an arbitrary odd integer). However, for $p>2$, fermion-parity is not enough to protect the $\mathcal{D}(\mathbb{Z}_p)$ neutral sector. For odd $p$, any $\mathbf{m}$-particle can be transformed into an $\mathbf{e}$-particle by adding/removing \textit{even} number of electrons. For even $p>2$, $\mathbf{m}^{2\mathbb{Z}}$-particles can be transformed into $\mathbf{e}^{2\mathbb{Z}}$-particles without changing fermion-parity. Therefore, except for $p=2$, both fermionic and bosonic non-diagonal states generally relies on the $U(1)$ charge symmetry to protect the $\mathcal{D}(\mathbb{Z}_p)$ neutral sector.

\subsubsection{Weak topological superconductor}\label{sec3.2.3}
The third example is related to the digression taken in Sec \ref{sec2.4.2}. There we have considered a coupled wire model of 2d weak topological superconductor (TSC), in which the vortex excitations have a similar constrained motion as the quasiparticles of non-diagonal quantum Hall states. Recall that the conservation of fermion-parity dictates the $h/2e$ vortex to be tunneled across two wires at a time, so the vortex excited on an even link should be differentiated from the one on an odd link. When the vortex is tunneled across two wires, notice that there is an accompanying tunneling of an electron between the wire, so braiding an $h/2e$ vortex on even links around another one on an odd link would require an electron to be transferred around a $\pi$-flux. This results in a braiding phase of $e^{i\pi}$. Equivalently, one could understand this by viewing the $\pi$-flux on an even link as a composite of a $\pi$-flux on an odd link together with a single electron. Therefore, the $\pi$-flux on even/odd links can be viewed as $\mathbf{e}/\mathbf{m}$-anyon in the $\mathbb{Z}_2$ toric code. This situation is similar to the $p=2$ fermionic non-diagonal state, in that none of them require $U(1)$ charge symmetry to protect the neutral sector. 

However, the weak TSC is different from the $p=2$ fermionic non-diagonal state in another important aspect: the $\mathbb{Z}$ translation symmetry in the weak TSC is \textit{not essential} for the $\mathbb{Z}_2$ topological order. While the translation symmetry acts as the $\mathbf{e} \leftrightarrow \mathbf{m}$ anyonic symmetry for the toric code (which is a virtue of the wire model), with or without this symmetry the superconductor always has a topological order. This is indeed a well-known fact: a fully-gapped superconductor coupled with \textit{dynamic electromagnetism} has a $\mathbb{Z}_2$ topological order \cite{Hansson2004SCisTO, Bonderson2013quasiTO, Radzihovsky2017TOinSC}. On the other hand, as we are going to elaborate below, the presence of translation symmetry is actually \textit{essential} to the $\mathcal{D}(\mathbb{Z}_p)$ neutral sector of non-diagonal QH states. Next, we discuss the importance of charge symmetry and translation symmetry in a more systematic manner.  

\subsection{Symmetry enrichment}\label{sec3.3}
We have now established that quasiparticles on the even/odd links can be associated to $\mathbf{e}$/$\mathbf{m}$-particles respectively. This leads us to interpret the non-diagonal states as having a $U(1)_{2pq}$ charge sector (for fermionic states it would be $U(1)_{p(p+2q)}$) and a $\mathcal{D}(\mathbb{Z}_p)$ neutral sector. This interpretation is useful as it consistently describes the fusion and braiding properties of the non-diagonal states. However, it is important to ask whether this interpretation is \textit{essential}. This is equivalent to asking whether the non-diagonal state is really different (if yes, then in what circumstances different) from a strongly-clustered state. From the perspective of a single wire that constitutes the wire model, as we have discussed in Sec. \ref{sec2.2}, these two states are respectively related to two distinct non-chiral CFTs, one at radius $R=\sqrt{p/2q}$ (\textit{i.e.} non-diagonal) while another at $R=1/\sqrt{2pq}$ (\textit{i.e.} diagonal), which suggests that the answer is yes. To fully answer the question we need to address the role of two symmetries: charge conservation and translation symmetry. The former has been hinted about in the examples just analyzed, while the latter is related to the $\mathbf{e}$-$\mathbf{m}$ anyonic symmetry. Here we discuss the symmetry issue from the bulk perspective, and in the next section we study the implications to the boundary. 

\subsubsection{Charge conservation}
From the specific examples analyzed in Sec.\ref{sec3.2}, we have seen that charge conservation plays an important role in constraining the quasiparticle scattering pattern. Here we provide a more general argument to establish the $U(1)$ charge symmetry as a necessary ingredient to protect the neutral sector. We focus on the bosonic states first. As discussed in Sec. \ref{sec2.3}, certain local scattering operators are charged for non-diagonal states ($p>1$), which indicate that hopping the associated quasiparticle across a single wire would violate charge conservation. Instead, the charge-conserving process is for a quasiparticle to hop across two wires at a time, thus differentiating excitations on the even and odd links. In the absence of $U(1)$ charge symmetry, however, such a distinction would be meaningless. For the bosonic non-diagonal state at filling $\nu=p/2q$, a local electron operator $e^{-i\varphi_j}$ creates an $(2q, \mathbf{e}^{-r_0q}\mathbf{m}^q)$-excitation, which is trivial from the perspective of intrinsic topological order. By fusing with $(2q, \mathbf{e}^{-r_0q}\mathbf{m}^q)$-excitations, all $\mathbf{m}$-particles can be transformed to $\mathbf{e}$-particles, thus rendering the neutral sector label meaningless. Hence, \textit{without} charge conservation the non-diagonal state of electron at filling $\nu=p/2q$ is topologically equivalent to the Laughlin state of $pe$-clusters at filling $\nu_{pe}=1/2pq$ (or the strongly-clustered state for short). In the presence of $U(1)$ charge symmetry, quasiparticles should be distinguished not only by their braiding statistics but also by their symmetry charge. This in turn distinguishes the $\mathbf{e}$-particles from the $\mathbf{m}$-particles in non-diagonal states. The only trivial quasiparticle, in the ``symmetry-enriched" sense, is $(0,\mathds{1})$. While it is also true for the strongly-clustered state that quasiparticles with different electric charge should be distinguished in the presence of $U(1)$ symmetry, there is no enriched neutral sector in that case.

The situation is similar for the fermionic states. Following Eq. (\ref{fermioniclocalop2maintext}), the local operator $\mathcal{O}^{\{-2,0\}}_j$ creates the $(2(p+2q), \mathbf{e}^{-2r_0q}\mathbf{m}^{2(p+q)})$-excitation. This is equivalent to a pair of electrons, so the fermion-parity is preserved. Notice that when $p$ is odd, $2(p+q)$ is coprime to $p$ (given our assumption that $p$ and $q$ are coprime), thus fusing with an appropriate number of the $(2(p+2q), \mathbf{e}^{-2r_0q}\mathbf{m}^{2(p+q)})$-excitations can turn any $\mathbf{m}$-particles into $\mathbf{e}$-particles. When $p$ is even, fusing with the $(2(p+2q), \mathbf{e}^{-2r_0q}\mathbf{m}^{2(p+q)})$-excitations would identify the $\mathbf{m}^{2\mathbb{Z}}$-particles with $\mathbf{e}^{2\mathbb{Z}}$-particles. Therefore, except for the $p=2$ states, a fermionic non-diagonal state also relies on the $U(1)$ charge symmetry to protect its $\mathcal{D}(\mathbb{Z}_p)$ neutral sector. 

Having said that, $U(1)$ charge symmetry is only necessary but not sufficient for distinguishing the non-diagonal states from the strongly-clustered state.

\subsubsection{Translation and anyonic symmetry}

In the presence of $U(1)$ charge conservation, excitations on the even links are distinguished from those on the odd links. However, without the translation symmetry that transforms wire $j \mapsto j+1$ (which we denote as the $\mathbb{Z}$ translation), the non-diagonal state is actually \textit{adiabatically connected} to the strongly-clustered state. This can be seen if we dimerize the $2j$-th wire with the $(2j+1)$-th wire (for all $j\in\mathbb{Z}$) such that the inter-wire couplings in Eq. (\ref{wireHamtot}), \textit{i.e.} $t_{2j+1/2}$'s, are pushed to infinity. This corresponds to setting the gap in the even links to be infinite, and thus all $\mathbf{e}$-particles are infinitely heavy and only $\mathbf{m}$-particles are left in the spectrum. In this way, the neutral sector becomes meaningless as the quasiparticle spectrum is the \textit{same} for both the non-diagonal state and the strongly-clustered state. In fact, they have the same topological ground state degeneracy on a torus: for bosonic state, the degeneracy is $N=2pq$; for fermionic state, the degeneracy is $N=p(p+2q)$. Thus, \textit{without} the $\mathbb{Z}$ translation symmetry, the non-diagonal state of electron at filling $\nu=p/2q$ is topologically equivalent to the Laughlin state of $pe$-bosons at filling $\nu_{pe}=1/2pq$. To sum up, the non-diagonal state is different from the strongly-clustered state precisely in that the former can be \textit{enriched} to a more exotic topological phase with an additional $\mathcal{D}(\mathbb{Z}_p)$ neutral sector, by the $U(1)\times \mathbb{Z}$ symmetry. 

It is important to notice that the $\mathbb{Z}$ translation symmetry in the wire model is related to the $\mathbb{Z}_2$ anyonic symmetry in the $\mathcal{D}(\mathbb{Z}_p)$ quantum double model, because the translation by one wire would transform even links to odd links, thus exchanging $\mathbf{e} \leftrightarrow \mathbf{m}$. In this regard, the non-diagonal QH state is similar to Kitaev's toric code on honeycomb lattice and Wen's plaquette model \cite{Kitaev2006exactly, You2012plaquette, You2013plaquette}, as well as the weak TSC discussed in Sec. \ref{sec3.2.3}. Nevertheless, the $\mathcal{D}(\mathbb{Z}_p)$ in a non-diagonal state is only realized with symmetry-enrichment, while the $\mathcal{D}(\mathbb{Z}_p)$ in Kitaev's and Wen's models (as well as the $\mathcal{D}(\mathbb{Z}_2)$ in superconductor) is intrinsic. 

We also want to comment on a subtle relation between the translation symmetry and the anyonic symmetry. When speaking of anyonic symmetry, it is usually viewed as an abstract \textit{relabeling} symmetry that permutes the anyon types in a way that all fusion and braiding properties are left invariant \cite{Teo2015twistliquid, Teo2016AS,Barkeshli2019AS}. In such a definition, no explicit reference to the Hamiltonian is made, and thus the energetics of anyons are really not concerned. In our case, it is then more precise to relate the translation symmetry to the \textit{exact} anyonic symmetry. The \textit{exactness} lies in the energetics, which requires the anyonic excitations to have the same energy under the relabeling transformation. 

\begin{figure}[t!]
   \includegraphics[width=7cm,height=5cm ]{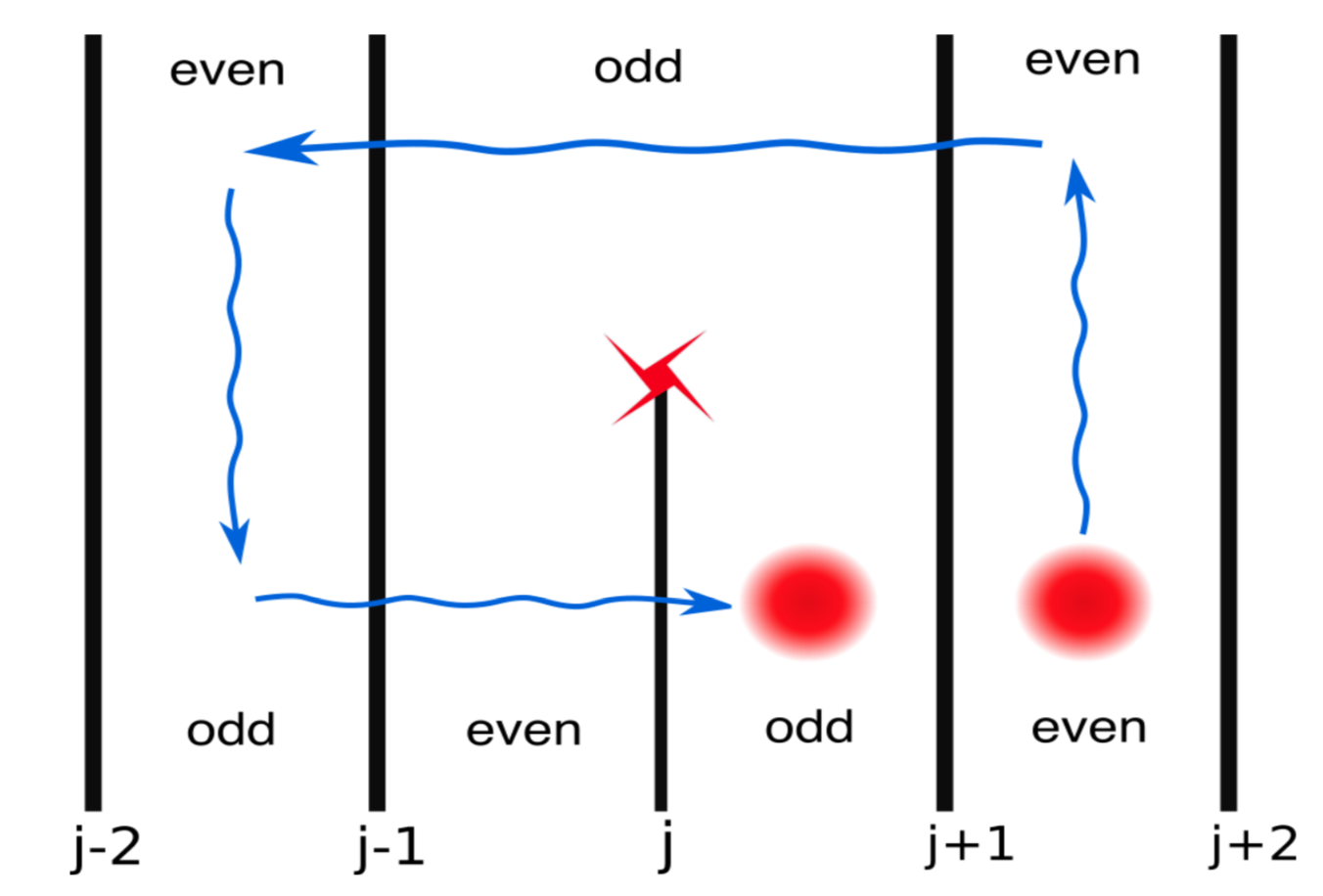}\centering
  \caption{\small{Illustration of a braiding process between a quasiparticle and a dislocation defect in the wire model, which turns a quasiparticle on the even link ($\mathbf{e}$-particle) into a quasiparticle on the odd link ($\mathbf{m}$-particle). A dislocation in the wire model thus acts as a twist defect for the anyonic symmetry in the $\mathcal{D}(\mathbb{Z}_p)$ quantum double model. }}
  \label{dislocation}
\end{figure}

The presence of an exact anyonic symmetry in the non-diagonal states has an important physical consequence, as then the symmetry can be gauged. According to the general theory of anyonic symmetry, the gauged phase is non-Abelian in nature \cite{Teo2015twistliquid, Teo2016AS, Barkeshli2019AS}. In the coupled wire model there is an explicit description of such gauging process, which is the proliferation of dislocation defects. A dislocation defect in the wire model is a sudden termination of a wire in the bulk, as illustrated in Fig. \ref{dislocation}.  Braiding an $\mathbf{e}$-particle around a dislocation defect would \textit{relabel} the quasiparticle as an $\mathbf{m}$-particle, and vice versa. Hence a single dislocation defect is a gauge flux of the $\mathbb{Z}_2$ anyonic symmetry, which is also known as a twist-defect \cite{Bombin2010AS}. To be more precise, as double-dislocations are condensed, a single dislocation would be truly $\mathbb{Z}_2$ in nature. In the ``melting phase" of the coupled wire model where single-dislocations are deconfined and double-dislocations proliferate, the $\mathbb{Z}_2$ anyonic symmetry in the neutral sector of the (Abelian) non-diagonal quantum Hall state is gauged, and the resulting phase would be isotropic and non-Abelian. We hope to better characterize this exotic quantum Hall state in future works. 

Now, equipped with the knowledge in the bulk we shall revisit the edge and discuss a signature of the symmetry-enrichment in non-diagonal states. When the anyonic symmetry in the bulk is exact, that is when the wire model has a discrete translation symmetry, an additional gapless boundary theory could emerge. 

\section{\label{sec4}Theory of the symmetric edge}
We now study the boundary theory in more detail. The coupled wire model has two types of edges: one is the left/right side edge which has been studied in Sec. \ref{sec2.2}; another type is the top/bottom edge that is formed by coupling together the ends of wires.  In this section, by studying the top/bottom edge, we first recover the chiral Luttinger liquid which has been shown to live on the left/right edge (Sec. \ref{sec4.1}). This is the chiral edge theory of the $U(1)_{2pq}$ charge sector. Without additional symmetry, it describes the \textit{only} gapless edge mode for the non-diagonal state, which is the same edge mode for a strongly-clustered state. This is reflecting that these states share the same intrinsic topological order, as explained in the previous section. However, with the $\mathbb{Z}$ translation symmetry in the wire model, a non-chiral gapless theory could emerge in the neutral sector, which describes the critical transition of a quantum $\mathbb{Z}_p$ clock model (Sec. \ref{sec4.2}). When both the charge and neutral sectors are gapless, a single electron can be tunneled into the symmetric edge from a metal. The associated tunneling exponent is predicted in Sec. \ref{sec4.3}, which may serve as a possible experimental signature for the non-diagonal states. 

What we discover for the symmetry-enriched neutral sector corroborates with earlier studies on critical parafermion chains \cite{criticalparafermionchain} and twist defect chains \cite{twistdefectchain}, which linked together translation invariance with self-duality of the clock model. Indeed, the symmetric edge of a non-diagonal quantum Hall state provides an electronic platform to realize the physics discussed in these earlier works. Given the discussion in Sec. \ref{sec2.4.2}, one would expect the ends of wires to host parafermion zero modes, which are coupled by electron-tunneling to form a parafermion chain at the edge. Alternatively, the discussion in Sec. \ref{sec3.3} suggests the termination of a wire as a twist defect of the $\mathbb{Z}_p$ toric code that exchanges $\mathbf{e}$ and $\mathbf{m}$ particles, so the top/bottom edge can be equivalently viewed as a twist defect chain. While in Ref. \cite{twistdefectchain} the equivalence between the twist defect chain and the clock model is demonstrated using Wilson loop operators, in our following analysis we intend to provide a more transparent derivation based on inter-wire electron-tunneling interactions at the edge. Importantly, we notice that a \textit{generalized} quantum clock model is actually realized at the edge, in contrast to the conventional clock model discussed previously. This complicates the situation for $p\geq 4$, and in Sec. \ref{sec4.4} we address the related subtleties.


For convenience, our discussions in Sec. \ref{sec4.1} and \ref{sec4.2} are based on the bosonic states. The results for the fermionic states are essentially the same, differ simply by a substitution $2q \mapsto p+2q$. In Sec. \ref{sec4.3}, where we discuss possible experimental signatures by tunneling electrons from Fermi liquid into the symmetric edge, we focus only on the fermionic states. 

\subsection{Charge sector}\label{sec4.1}
\begin{figure}[b!]
   \includegraphics[width=8cm,height=5.5cm ]{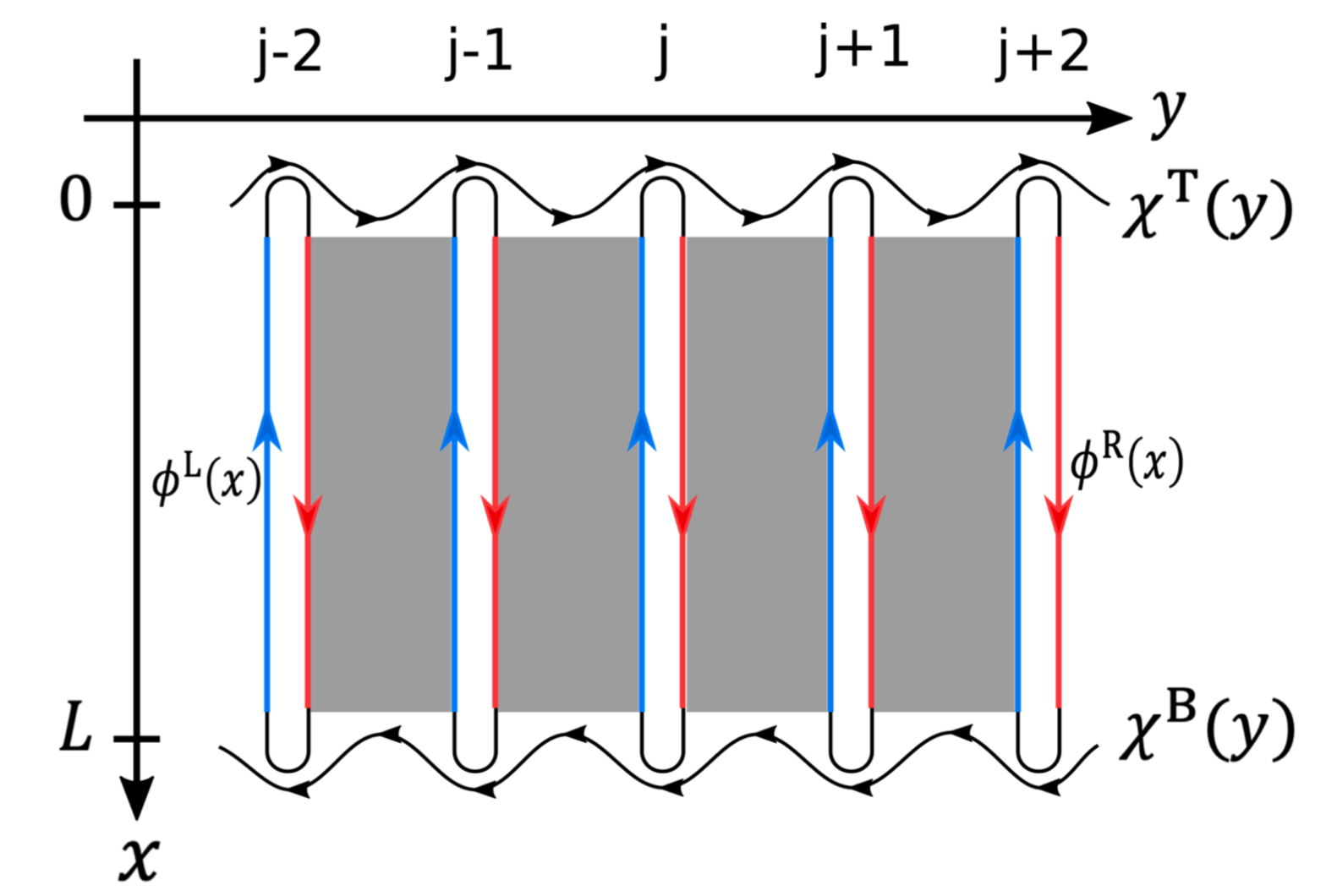}\centering
  \caption{\small{Chiral Luttinger liquids at the top (T) and bottom (B) edges of the coupled wire model, labeled as $\chi^\text{T}$ and $\chi^\text{B}$ respectively. The grey shaded region represents the gapped bulk, obtained from inter-wire tunneling of charge-$pe$ clusters. The termination of each wire is modeled by a hard-wall boundary condition, such that the chiral and anti-chiral modes of each wire ($\phi^\text{R}$ and $\phi^\text{L}$) are reflected into each other. While $\phi^{\text{R}/\text{L}}$ on neighboring wires are locked together deep in the bulk, they are left to fluctuate near the boundary where the inter-wire couplings vanish, giving rise to the gapless charge mode $\chi^{\text{T}/\text{B}}$.}}
  \label{edgecharge}
\end{figure}

The edge theory in the charge sector can be intuitively understood in a pictorial depiction of the coupled wire model as shown in Fig. \ref{edgecharge}. While the inter-wire couplings have gapped out the bulk by freezing the degrees of freedom therein, a chiral Luttinger liquid is left freely fluctuating near the termination of wires, where the inter-wire couplings diminish. Here we provide a more rigorous derivation of this Luttinger liquid edge mode, and the setup would also be useful for understanding the more non-trivial edge modes in the neutral sector. 

To model the termination of a wire, we adopt the hard-wall boundary condition so that left-movers are reflected into right movers, and vice versa at the other end. The finite-size Luttinger liquid is then characterized by the bosonized variables $\varphi(x)$ and $\theta(x)$ that satisfy
\begin{equation}\label{hardwallcommutation}
[\theta_j(x), \varphi_{j'}(x')]= i\pi \delta_{jj'} H(x-x'), 
\end{equation}
where $H(x)$ is the Heaviside step function, together with the boundary conditions
\begin{equation}
\theta_j(0^-) =  0\;\;\text{and}\;\;\theta_j(L^+)=\pi N_j. 
\end{equation}
Here $j,j'$ label the wires, which terminate at $x=0,L$, and $N_j$ is the electron number operator for wire $j$. Importantly, we have been careful in specifying the $x$-coordinates of the bosonic fields, in the above and in what follows, so as to ensure that commutation relations can always be evaluated unambiguously. In our notation, the very end of the wire with a fixed boundary condition is located at $x=0^-\;(L^+)$, the inter-wire tunneling that fluctuates near the boundary happens at $x=0\;(L)$, and the inter-wire tunneling that is pinned in the bulk is thought of as happening at $x=0^+\;(L^-)$. This seemingly pedantic effort would prove to be crucial when we derive the chiral algebra for the top/bottom edge mode. 

 In terms of the chiral modes introduced in Eq. (\ref{chiralboson}), we have 
\begin{subequations}
\begin{align}
[\phi^{\text{R}/\text{L}}_j(x), \phi^{\text{R}/\text{L}}_{j'}(x')] &= \pm 2i\pi pq \delta_{jj'}\text{sgn}(x-x'),\\
[\phi^{\text{R}}_j(x), \phi^{\text{L}}_{j'}(x')] &= 2i\pi pq \delta_{jj'}, 
\end{align}
\end{subequations}
together with the boundary conditions
\begin{subequations}
\begin{align}
&\phi^\text{R}_j(0^-) = \phi^\text{L}_j(0^-),\\
&\phi^\text{R}_j(L^+)= \phi^\text{L}_j(L^+)+4\pi qN_j.
\end{align}
\end{subequations}
Notice that there are generally discontinuities in these chiral modes at the edge ($x=0, L$) from one wire to the next, which are caused by the inter-wire tunneling term $\cos\Theta_{j+1/2}$. Indeed, $\phi^{\text{R}/\text{L}}(x)$ is the chiral mode of each single wire defined along the $x$-direction, so they are not quite the right variables for describing the top/bottom edge modes which run along the $y$-direction. 

To identify the appropriate chiral edge modes at $x=0\; ({\rm top})$ and $x= L\;({\rm bottom})$, which should vary slowly from one wire to the next, let us examine again the bulk inter-wire coupling, but now slightly modified to $\cos\tilde{\Theta}_{\ell}(x)$, with the link variable
\begin{equation}
\tilde{\Theta}_{j+1/2}(x) \equiv \phi^\text{R}_j(x)-\phi^\text{L}_{j+1}(x)-2\pi q N_j.
\end{equation}
Compared with Eq. (\ref{link1}), the inter-wire coupling is defined with an extra $2\pi q N_j$ term. This modification is needed to ensure that $[\tilde{\Theta}_{\ell}(x),\tilde{\Theta}_{\ell'}(x')]=0$, given the commutation relation in Eq. (\ref{hardwallcommutation}) which is appropriate for a hard-wall boundary condition. As we have shown in Sec. {\ref{sec2.1}, the inter-wire couplings then pin $\tilde{\Theta}_{\ell} \in 2\pi \mathbb{Z}$ everywhere in the bulk, and thus completely gap out the bulk. At the boundaries $(x=0,L)$, the inter-wire interaction diminishes so that $\tilde{\Theta}_{\ell}$ is allowed to fluctuate there. As we see next, this fluctuation gives rise to the chiral Luttinger liquid at the top/bottom edge.

We now introduce the chiral edge mode living at top/bottom ($x=0/L$) edge as follows,
\begin{subequations}\label{chargemode}
\begin{align}
\chi_j(0)  &= \phi^\text{L}_j(0)-2\pi q \sum_{j \leq i} N_i -\sum_{j\leq i}\tilde{\Theta}_{i+1/2}(0^+),\\
\chi_j(L)  &= \phi^{\text{L}}_j(L)+2\pi q \sum_{j \leq i} N_i -\sum_{j\leq i}\tilde{\Theta}_{i+1/2}(L^-), 
\end{align} 
\end{subequations}
where $\tilde{\Theta}_{\ell}(0^+)$ and $\tilde{\Theta}_{\ell}(L^-)$ correspond to the bulk link variables that are pinned. For link $\ell=j+1/2$, the link variables at the edge are then
\begin{subequations}
\begin{align}
\tilde{\Theta}_{\ell}(0) &= \chi_j(0)-\chi_{j+1}(0)+\tilde{\Theta}_{\ell}(0^+),\\
\tilde{\Theta}_{\ell}(L) &= \chi_j(L)-\chi_{j+1}(L)+\tilde{\Theta}_{\ell}(L^-), 
\end{align}
\end{subequations}
which imply that $\chi_j(0/L)$ indeed varies slowly between neighboring wires. The fluctuation of $\chi$ is controlled by the inter-wire tunneling near the boundary, which is proportional to
\begin{equation}
\cos \tilde{\Theta}_{\ell}(0/L) \sim (\chi_j(0/L)-\chi_{j+1}(0/L))^2.
\end{equation}
Note that the series expansion is legitimate because $\tilde{\Theta}_{\ell}(0^+)$ and $\tilde{\Theta}_{\ell}(L^-)$ are pinned at $2\pi\mathbb{Z}$.   Taking the continuum limit in the $y$-direction, \textit{i.e.} $\chi_j(0) \mapsto \chi^\text{T}(y)$ and $\chi_j(L) \mapsto \chi^\text{B}(y)$, we obtain the effective Hamiltonian for the top/bottom edge,
\begin{equation}\label{top/bottomHam}
\mathcal{H}^{\text{T}/\text{B}}_{\rho}= \frac{u}{2\pi}(\partial_y\chi^{\text{T}/\text{B}})^2.
\end{equation}
Furthermore, one can readily check that 
\begin{subequations}
\begin{align}
[\chi_j(0),\chi_{j'}(0)] &= 2i\pi pq\;\text{sgn}(j-j'),\\
[\chi_j(L),\chi_{j'}(L)] &= -2i\pi pq\; \text{sgn}(j-j'), 
\end{align}
\end{subequations}
which imply the chiral algebra in the continuum limit, 
\begin{equation}\label{top/bottomchiralalgebra}
[\chi^{\text{T}/\text{B}}(y),\chi^{\text{T}/\text{B}}(y')] = \pm\;2i\pi pq\;\text{sgn}(y-y').
\end{equation}
Altogether, Eqs. (\ref{top/bottomHam}) and (\ref{top/bottomchiralalgebra}) suggest that the low-energy effective theory for the top/bottom edge of the $\nu=p/2q$ non-diagonal state is \textit{partly} described by a chiral Luttinger liquid with Luttinger parameter $K=2pq$. A similar result holds for the fermionic state at filling $\nu=p/(p+2q)$, with $K=p(p+2q)$. This is the edge mode guaranteed by the bulk topological order, and it coincides with the gapless mode on the left/right side edge described by $\phi^{\text{R}/\text{L}}$. The subscript $\rho$ in Eq. (\ref{top/bottomHam}) represents the charge sector, and as we discuss next, the edge Hamiltonian could have other contributions that would be attributed to the neutral sector $(\sigma)$, which become particularly important in the presence of symmetry. 

\subsection{Neutral sector}\label{sec4.2}

\subsubsection{Physical picture}\label{sec4.2.1}
In the bulk of non-diagonal states, wires of Luttinger liquid are coupled together by inter-wire tunneling of $p$ electrons. As shown in Sec. \ref{sec2.1}, at electron filling $\nu=p/2q$ the bulk is completely gapped, so the $pe$-tunneling is the only interaction that matters in the bulk. This leaves a gapless chiral Luttinger liquid fluctuating at the boundary as we have shown above. This interaction preserves the electron number mod $p$ in each wire. From now on this quantity is referred to as the ``number $p$-rity".

Note, however, the number $p$-rity of each wire is generally \textit{not} conserved. By tunneling a single electron between the ends of two neighboring wires, e.g. $e^{i(\varphi_{j}-\varphi_{j+1})}$, the number $p$-rity of each involved wire is shifted by $1$. Given that the charge sector at the boundary (associated to $pe$-tunneling) is gapless, the inter-wire tunneling of a single electron could be important at the boundary. Thus, a complete description of the edge should take into account all possible fluctuations of the number $p$-rity of each wire. To gain physical insights, say for the top edge, we pretend to dimerize the array of wires by connecting wire $2j$ with wire $2j+1$ (for all $j \in \mathbb{Z}$) at the bottom edge, as depicted in Fig. \ref{edgeneutral}. The $x=0$ end of wire $2j$ and the $x=0$ end of wire $2j+1$ then become two ends of the \textit{same} Luttinger liquid, and the electron tunneling between them, \textit{i.e.} $e^{i(\varphi_{2j}-\varphi_{2j+1})}$, would conserve the number $p$-rity of this Luttinger liquid. We expect the inter-wire tunneling over link $2j+1/2$ to be related to a $p$-state clock operator $\tau_j$ that measures this number $p$-rity, while the inter-wire tunneling over neighboring links ($2j-1/2$ and $2j+3/2$) to be related to a shift operator $\sigma_j$ that changes this number $p$-rity. The effective Hamiltonian then describes a $p$-state clock model. 

\begin{figure}[t!]
   \includegraphics[width=8cm,height=5.5cm ]{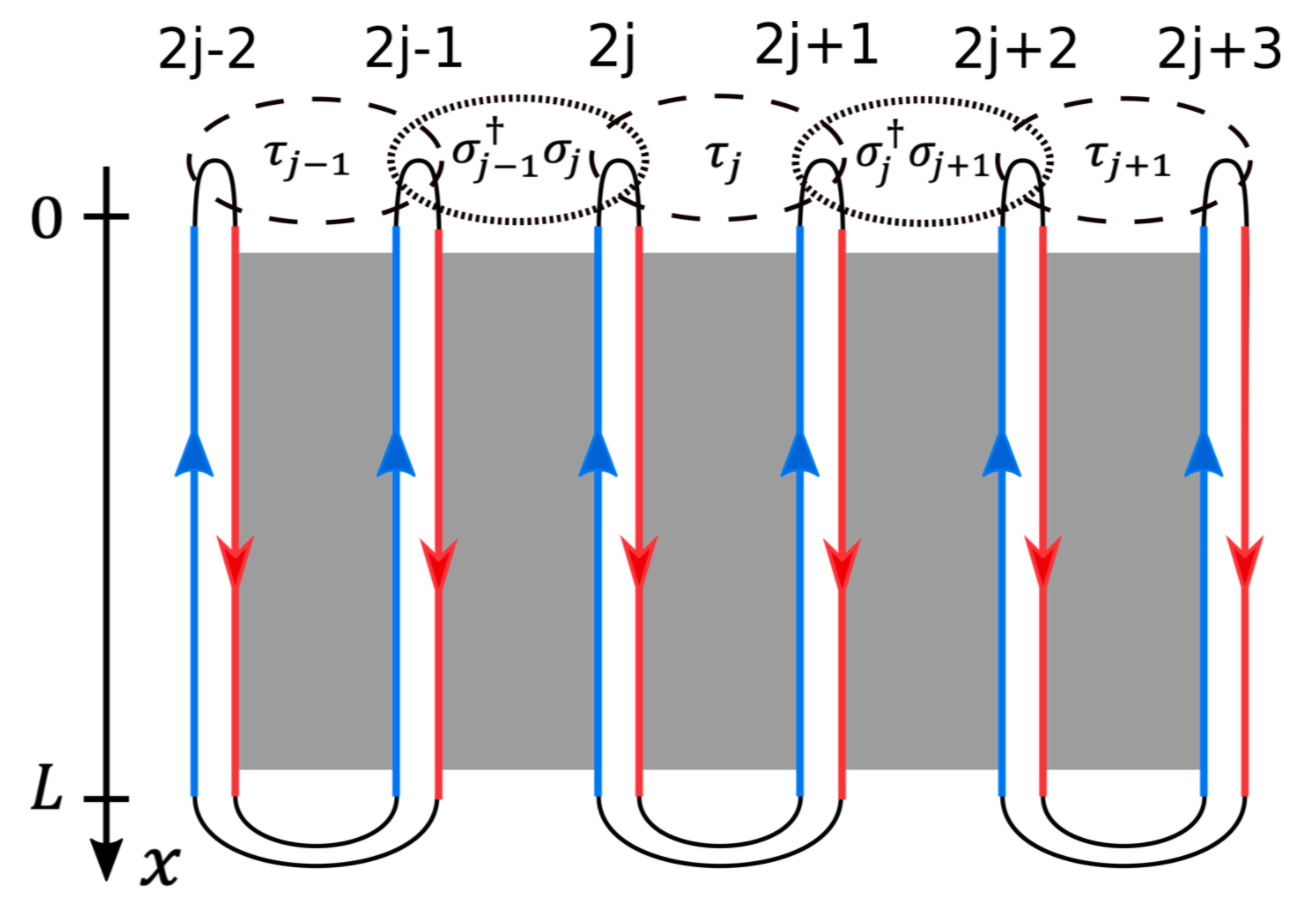}\centering
  \caption{\small{Quantum $\mathbb{Z}_p$ clock model at the top edge of the coupled wire model, which is obtained from inter-wire tunneling of a single electron. As an aid of thinking, we imagine that the array of wires are dimerized such that wire $2j$ and wire $2j+1$ are connected at the bottom edge, forming a single Luttinger liquid, to which we associate a number $p$-rity. Tunneling between wire $2j$ and $2j+1$ (dashed circle) preserves this number $p$-rity, while tunneling between wire $2j$ and $2j-1$ (dotted circle), as well as between wire $2j+1$ and $2j+2$, shift the number $p$-rity. These edge-coupling can be associated to the clock operators $\tau_j$ and shift operators $\sigma_j$ as shown in the main text. Choosing a different dimerization pattern is equivalent to an order-disorder duality transformation. Given the translation symmetry in the bulk of wire model, the clock model at the edge is self-dual.}}
  \label{edgeneutral}
\end{figure}

The dimerization procedure just described is fictitious, but it provides an intuitive perspective for understanding the edge neutral sector. In particular, it naturally leads to the order-disorder duality in the clock model. Had we chosen another dimerization pattern, which connects wire $2j$ with wire $2j-1$, we would have associated a dual clock operator $\nu_{j-1/2}$ to measure the number $p$-rity over link $2j-1/2$, and a dual shift operator $\mu_{j-1/2}$ to change this number $p$-rity. Importantly, when the $\mathbb{Z}$ translation symmetry is present in our coupled wire model, dimerization is actually \textit{forbidden}. The two ways of dimerization described above are thus put on the same footing, which suggests that the edge neutral sector is a \textit{self-dual} clock model, described by some gapless critical theory in the continuum limit. Next, we will supplement the above argument by a more rigorous derivation. We focus on the $x=0$ (top) edge, as the situation for the bottom edge is essentially the same.

\subsubsection{Generalized $\mathbb{Z}_p$ clock chain}\label{sec4.2.2}
Let us first consider the inter-wire tunneling of a single electron,
\begin{equation}
H_{1e} = -J_1 \sum_{j} \cos(\varphi_{j}-\varphi_{j+1}).
\end{equation}
The notation at the boundary is simplified, \textit{i.e.} $\varphi_j \equiv \varphi_j (0)$. We also assume translation symmetry here, so that the tunneling strength is the same for each link. Later on, we will discuss the physical consequences with/without this symmetry. Using Eq. (\ref{chargemode}), we have
\begin{equation}\label{singletunnelterm}
\varphi_j-\varphi_{j+1} = \frac{1}{p}(\chi_j-\chi_{j+1})+\frac{2\pi q}{p}N_j+\frac{2\pi}{p}\widetilde{N}_{j+1/2}.
\end{equation}
For later convenience, we have introduced the quasiparticle number operator $\widetilde{N}_{\ell} = \tilde{\Theta}_{\ell}(0^+)/2\pi$. By definition, $\widetilde{N}_{\ell}$ has integer eigenvalue, and it is shifted by $1$ whenever a minimal quasiparticle is tunneled from one end of the link to another. The first term in Eq. (\ref{singletunnelterm}), which involves $\chi_j-\chi_{j+1}$, simply contributes to the Luttinger liquid in the charge sector. The remaining terms represent additional contributions in the neutral sector that we are interested in. This motivates us to introduce the following operator,
\begin{equation}
\mathcal{W}_{j+1/2}  = e^{i(\frac{2\pi q}{p}N_j+\frac{2\pi}{p}\widetilde{N}_{j+1/2}+\pi q)}.
\end{equation}
One can readily check that $\mathcal{W}_{\ell}^p=1$, which follows from the commutation relation
\begin{equation}
[N_j, \widetilde{N}_{k+1/2}] = \frac{ip}{2\pi}(\delta_{j,k}-\delta_{j,k+1}).
\end{equation}
Moreover, these operators satisfy the commutation algebra appropriate for a quantum $\mathbb{Z}_p$ clock model,
\begin{subequations}
\begin{align}
[\mathcal{W}_{j+1/2}, \mathcal{W}_{k+1/2}] = 0,\; \text{for }\abs{j-k}>1, \\
\mathcal{W}_{j+1/2}\mathcal{W}_{j-1/2}=\omega\;\mathcal{W}_{j-1/2}\mathcal{W}_{j+1/2},
\end{align}
\end{subequations}
with $\omega=e^{2\pi i q/p}$ \cite{clockfermionic}. This reflects the physical intuition we discussed earlier: the single-electron tunneling through each link shall be associated to a $p$-state clock operator, while the tunneling through the neighboring link shall be treated as the corresponding shift operator. We can make an explicit correspondence to the $\mathbb{Z}_p$ clock model by defining the clock variables as follows,
\begin{subequations}\label{topclockvar}
\begin{align}
\mathcal{W}_{2j+1/2} &= \tau_j,\\
\mathcal{W}_{2j-1/2} &= \sigma_j\sigma^\dagger_{j-1}.
\end{align}
\end{subequations}
They satisfy
\begin{subequations}\label{topclockalg}
\begin{align}
&\tau_j^p = \sigma_j^p =1,\\
&\tau_j \sigma_j = \omega \sigma_j \tau_j, 
\end{align}
\end{subequations}
and $\tau_j$ commutes with $\sigma_{k}$ for $j\neq k$. Consequently, the Hamiltonian for the inter-wire tunneling of a single electron can be written as,
\begin{equation}
H_{1e} = -J_1\sum_j (\tau_j+\sigma_j\sigma^\dagger_{j-1})+\text{H.c.}\;, 
\end{equation}
Only the contribution in the neutral sector is considered here, as the charge sector has been taken into account already. 

More generally, one should consider all possible $ne$-tunneling processes, for $ 1\leq n < p$. The effective Hamiltonian in the neutral sector thus takes the following form,
\begin{equation}\label{genclockHamiltonian}
H_\sigma = -\sum_j \sum_{n=1}^{p-1} J_n[(\tau_j)^n +(\sigma_j\sigma^\dagger_{j-1})^n+\text{H.c.}]\;.
\end{equation}
Importantly, the spin-spin coupling and the transverse-field coupling have the same strength due to the translation symmetry which interchanges even and odd links. As we have argued with a physical picture, translation in the bulk (by a single wire) actually associates to the order-disorder duality at the edge. Indeed, we can introduce the dual clock variables as follows,
\begin{subequations}\label{topdualclock}
\begin{align}
\mu_{j-\frac{1}{2}} &= \prod_{j \leq i} \tau^\dagger_i,\\
\nu_{j-\frac{1}{2}} &= \sigma_{j} \sigma^\dagger_{j-1},
\end{align}
\end{subequations}
which is a ``$p$-state'' generalization to the Kramers-Wannier duality transformation for the Ising model \cite{KW1941, Onsager1944, Kaufman1949}. The Hamiltonian can then be rewritten as
\begin{equation}
H_\sigma = -\sum_j \sum_{n=1}^{p-1} J_n[(\nu_{j-\frac{1}{2}})^n +(\mu_{j+\frac{1}{2}}\mu^\dagger_{j-\frac{1}{2}})^n+\text{H.c.}]\;.
\end{equation}
This demonstrates the self-duality of the $p$-state clock model, provided that the wire model has translation symmetry, which in turn is equivalent to the bulk $\mathbf{e}$-$\mathbf{m}$ anyonic symmetry according to our discussion in Sec. \ref{sec3.3}. 

Some words of caution are due here. We would refer to the symmetry-enriched neutral sector Hamiltonian in Eq. (\ref{genclockHamiltonian}) as the self-dual \textit{generalized} $\mathbb{Z}_p$ clock model, which is to be contrasted with the conventional clock model for $p\geq 4$. In 2d classical statistical mechanics, the distinction between the general clock model and the conventional one (see \cite{Elitzur1979pureclock}) is discussed by Cardy in Ref. \cite{Cardy1980generalclock}. Here for the 1d quanutm chain, the differences are two-fold: firstly, the clock operators are defined to obey Eq. (\ref{topclockalg}) with $\omega=e^{2\pi i q/p}$, in contrast to $\omega=e^{2\pi i/p}$ in the conventional model. Our model thus have distinct self-dualities for different $q$'s; secondly, the clock operators here appear with various powers, \textit{i.e.} $(\tau_j)^n$ and $(\sigma_j\sigma^\dagger_{j-1})^n$ with $n$ ranging from $1$ to $p-1$, in contrast to the conventional model with just $n=1$ \cite{Ortiz2012pureclock,Ortiz2019pureclock}. Consequently, for $p\geq 4$ the generalized model is different form the conventional one. One has to pay special attention to the more complicated phase diagram at self-duality \cite{Alcaraz1980generalclock, Dorey1996generalclock}. As we are going to discuss in Sec. \ref{sec4.4}, the symmetry-enriched edge neutral sector can sometime be gapped. 

For $p=2,3$, the generalized clock model is no differnt from the conventional one. For $p=2$ the neutral sector is described by an Ising-Majorana chain \cite{Kogut1979, gogolin2004}, while for $p=3$ it is described by a three-state Potts chain \cite{Fendley2012parafermionzeromode, mong2014parafermionic}. Self-duality then implies a critical transition characterized by some gapless continuum theory. As is well-known in statistical mechanics, the corresponding gapless theories are the Ising CFT and the $\mathbb{Z}_3$ parafermion CFT respectively \cite{Ginsparg88, BYB, Fateev:1985mm, lecheminant2002criticality}. With both the charge and neutral sectors being gapless, a single electron can be tunneled into the symmetric edge. Such tunneling experiments may be used to probe the non-diagonal states. Our next task is to compute the edge tunneling exponents for non-diagonal states, especially for $p=2,3$, which have symmetry-protected gapless edges. 

\subsubsection{Edge operators}\label{sec4.2.3}

To that end, it is useful to express the edge electron operator $\psi_j(0) \propto e^{i\varphi_j(0)}$ in terms of operators in the charge and neutral sectors explicitly. To do so, let us define the lattice parafermion operator in the neutral sector by combining the order and disorder operators \cite{FradkinKadanoff1980},
\begin{subequations}
\begin{align}
&\beta_{2j} = \omega^{\frac{p-1}{2}} \mu^\dagger_{j-\frac{1}{2}}\sigma^\dagger_{j},\\
&\beta_{2j-1} = \mu^\dagger_{j-\frac{1}{2}}\sigma^\dagger_{j-1}, 
\end{align}
\end{subequations}
which satisfy $\beta_j^p=1$ and
\begin{equation}
\beta_j \beta_{k}  = \omega^{\text{sgn}(j-k)}\beta_{k}\beta_j, 
\end{equation}
where $\omega= e^{2i\pi q/p}$. Maneuvering through the definition of variables introduced in this section, one can verify that the edge electron operator can be expressed simply as follows,
\begin{equation}\label{latticeedgeelectron}
\psi_{j}(0) \propto \beta_j e^{\frac{i}{p}\chi_{j}(0)}.
\end{equation}
Therefore, in the continuum limit, the scaling dimension of the edge electron is
\begin{equation}\label{scalingofelectron}
\begin{split}
\Delta_e = \Delta_{\beta}+\frac{K}{2p^2} = \Delta_{\beta}+\frac{1}{2\nu}.
\end{split}
\end{equation}
Here, $\Delta_\beta$ is the scaling dimension of the (most relevant) continuum field corresponding to the lattice parafermion operator. The Luttinger parameter is $K=2pq$ for a bosonic state at filling $\nu=p/2q$, and $K=p(p+2q)$ for a fermionic state at $\nu=p/(p+2q)$. The above expression holds up as long as the charge and neutral sectors decouple at low energy. As we will explain, this is indeed the case for $p=2,3$.

The above discussion allows one to experimentally reveal the symmetry-enriched edge structure through the tunneling exponent for tunneling electrons from an ordinary metal into the symmetric edge. We will elaborate on this in the next subsection. Alternatively, one can consider the inter-edge quasiparticle tunneling through a point contact, which make use of the operators that scatter a minimal quasiparticle from the top edge to the bottom edge. For instance, for the bosonic states, one can check that $[\widetilde{N}_{j+1/2}, (\phi^\text{R}_j(0)-\phi^\text{R}_j(L))/2pq]=i$, hence the following operator tunnels a minimal quasiparticle of charge $e/2q$ from the top edge to the bottom edge through link $\ell=j+1/2$,
\begin{equation}
\Pi^{e/2q}_{\ell}=e^{\frac{i}{2pq}(\phi^\text{R}_j(0)-\phi^\text{R}_j(L))}.
\end{equation}
Combining the above discussions for both the charge and neutral sectors, we can re-express the inter-edge tunneling operator as
\begin{subequations}\label{interedgeqp}
\begin{align}
\Pi^{e/2q}_{ 2j+\frac{1}{2}} & \propto (\sigma^\text{T}_j)^{-r_0} (\sigma^\text{B}_j)^{r_0} e^{\frac{i}{2pq}(\chi_{2j}(0)-\chi_{2j}(L))}, \\
\Pi^{e/2q}_{2j-\frac{1}{2}} & \propto (\mu^{\text{T}}_{j-\frac{1}{2}})^{-r_0} (\mu^{\text{B}}_{j-\frac{1}{2}})^{r_0} e^{\frac{i}{2pq}(\chi_{2j}(0)-\chi_{2j}(L))}, 
\end{align}
\end{subequations}
where $r_0$ satisfies $qr_0 = 1 \text{ mod }p$. Here $\sigma^\text{T}$ and $\mu^\text{T}$ are the clock variables for the top edge defined in Eqs. (\ref{topclockvar}) and (\ref{topdualclock}), while $\sigma^\text{B}$ and $\mu^\text{B}$ are the clock variables for the bottom edge which can be defined analogously. The above expression suggests that quasiparticles excited on the even links, which are known as the $\mathbf{e}$-particles in Sec. \ref{sec3}, are created at the top/bottom edge with the spin operator $\sigma^{\text{T}/\text{B}}$. On the other hand, quasiparticles excited on the odd links, which are known as the $\mathbf{m}$-particles, are created with the disorder operator $\mu^{\text{T}/\text{B}}$. Again, we are seeing here the equivalence between the order-disorder duality at the edge and the $\mathbf{e}$-$\mathbf{m}$ anyonic symmetry in the bulk \cite{Lichtman2020bulkedge}. 

In principle, one could use Eq. (\ref{interedgeqp}) to compute the tunneling exponent for inter-edge quasiparticle tunneling at a point contact and thus reveal the structure of the symmetric edge. Having said that, in making the constriction, translation symmetry on the edge may be easily broken to render a gapped neutral sector. A more practical way of probing the symmetric edge structure is by tunneling electrons into the edge from a Fermi liquid, which is what we focus on in the following. For experimental relevance, we only consider the fermionic non-diagonal states.

\subsection{Tunneling from metal into the symmetric edge}\label{sec4.3}
In the presence of translation symmetry, both the charge and neutral sectors of the top/bottom edge are gapless for non-diagonal states with $p=2,3$. A single electron can then be tunneled into the symmetric edge. For the left/right side edge, however, the neutral sector is gapped and this edge is completely characterized by the chiral Luttinger liquid of charge-$pe$ clusters, which is gapless only to the tunneling of $p$ electrons. This anisotropy between the top/bottom and the left/right edges highlights the symmetry-enrichment aspect of the non-diagonal quantum Hall states. Experimentally, the edge structure of a quantum Hall state can be revealed by measuring the tunneling exponents \cite{kane1996edge, chang2003CLLedge}. In the following, we are mainly interested in the tunneling from an ordinary metal, a Fermi liquid, into the edge of fermionic non-diagonal state at filling $\nu=p/(p+2q)$. 


Before analyzing the symmetric edge, let us make a contrast with the situation where the translation symmetry is broken. In this case the clock model is no longer self-dual, so the neutral sector is generally gapped. The top/bottom edge is then identical to the left/right side edge. Both are described only by a chiral Luttinger liquid with $K=p(p+2q)$. Notice that, neither $e^{\frac{i}{p}\chi}$ nor $e^{\frac{i}{p}\phi}$ are local operators, hence a single electron cannot be tunneled into these edges. The most relevant local operator at the edge is either $e^{i\phi}$ or $e^{i\chi}$, which corresponds a charge-$pe$ cluster with scaling dimension $\Delta_{pe} = K/2$. The charge-$pe$ cluster in the Fermi liquid has scaling dimension $\delta_{pe}=p^2/2$. Thus, for the non-symmetric edge, the tunneling current $I$ has the following scaling \cite{chang2003CLLedge},
\begin{equation}\label{sideedgetunnel}
I \sim V^{2(\Delta_{pe}+\delta_{pe})-1} = V^{p^2/\nu+p^2-1},
\end{equation}
where $V$ is the bias voltage. The same tunneling exponent is obtained by tunneling from metal into the strongly-clustered state (Laughlin state of $pe$-clusters) at filling $\nu_{pe}=\nu/p^2=1/p(p+2q)$. This is expected given our discussion in Sec. \ref{sec3.3}: the non-diagonal state shares the same intrinsic topological order as a strongly-clustered state.

On the other hand, the symmetric edge is gapless to a single electron, at least for $p=2,3$, and this can be used to reveal the signature of symmetry-enrichment in the non-diagonal state. The tunneling current from metal into the symmetric edge has the following power-law behavior,
\begin{equation}\label{tunnelingexponentsymmetricedge}
I \sim V^{2(\Delta_e+\delta_e)-1}, 
\end{equation}
where $\delta_e=1/2$ and $\Delta_e$ is given by Eq. (\ref{scalingofelectron}) provided that charge and neutral sectors decouple. Let us now analyze the specific cases in detail.

\subsubsection{$p = 2$: Ising CFT}
We first note that the $U(1)$ charge sector decouples with the Ising neutral sector at low energy. To couple together the two sectors, one would consider an operator $\widehat{O}_{cn} = \widehat{O}_c \widehat{O}_n$, where $\widehat{O}_c$ and $\widehat{O}_n$ are local operators in the charge and neutral sectors respectively. In the charge sector, the most relevant non-trivial operator is $\partial_y \chi$, with scaling dimension 1. In the neutral sector, the spin field $\sigma$ is not local. In fact, as we have seen in last subsection, the spin operator $\sigma$ and the disorder operator $\mu$ correspond to the bulk anyons $\mathbf{e}$ and $\mathbf{m}$ respectively. As for the energy operator $\epsilon \sim \beta \bar{\beta}$ (with scaling dimension 1), while being local, it dimerizes the Ising spin chain and violates the translation symmetry. Therefore, the dominant allowed coupling is $\widehat{O}_{cn}=(\partial_y\chi)\mathcal{T}$, with $\mathcal{T}$ being the stress-energy tensor in the Ising CFT. The total scaling dimension of the coupling is $3$, hence irrelevant, which implies the decoupling between the charge and neutral sectors.

It then follows from Eq. (\ref{tunnelingexponentsymmetricedge}) that the edge tunneling current scales with the bias voltage as
\begin{equation}
I \sim V^{1/\nu+1},
\end{equation}
for the fermionic non-diagonal state at filling $\nu=1/(q+1)$, with $q \in 2\mathbb{Z}+1$. Here we have used $\Delta_\beta=1/2$ for the Majorana field \citep{BYB}. 

\subsubsection{$p = 3$: $\mathbb{Z}_3$ parafermion CFT}
The situation for $p=3$ is similar to $p=2$. For an operator $\widehat{O}_{cn} = \widehat{O}_c \widehat{O}_n$ coupling the charge and neutral sectors, $\widehat{O}_n$ again cannot be the spin or disorder operator as they are associated to creating the non-local $\mathbf{e}$/$\mathbf{m}$ quasiparticles in the bulk. Also, the translation symmetry at the edge forbids $\widehat{O}_n$ to be the energy operator with dimension $4/5$. The most relevant allowed coupling is then given by $\hat{O}_{cn}=(\partial_y\chi)\mathcal{T}$, where $\mathcal{T}$ is the stress tensor for the $\mathbb{Z}_3$ parafermion CFT. Again, with scaling dimension 3, this coupling is irrelevant at low energy. Hence, the $U(1)$ charge sector and the $\mathbb{Z}_3$ parafermion neutral sector are decoupled at infra-red on the symmetric edge.

The $\mathbb{Z}_3$ parafermion is little more subtle than the Majorana fermion, as the continuum limit of the lattice parafermion operator is not just the parafermion primary field. As argued by Mong \textit{et al.} \cite{mong2014parafermionic}, aside from the parafermion field with dimension $2/3$, the lattice parafermion operator actually contains a more relevant primary field with scaling dimension $7/15$. Thus, we should use $\Delta_\beta=7/15$ for $p=3$. This leads to the following scaling relation between the tunneling current and the bias voltage,
\begin{equation}
I \sim V^{1/\nu+14/15},
\end{equation}
for the fermionic non-diagonal state at filling $\nu=3/(2q+3)$, with $q \in 3\mathbb{Z}\pm 1$.

\subsection{Complexities for $p \geq 4$: the generalized clock model}\label{sec4.4}

Finally, let us comment on the edge structure of non-diagonal states with $p\geq 4$. Unlike cases for $p<4$, translation symmetry (or self-duality) alone does not guarantee a gapless neutral sector. Our following discussion supplements the results obtained in Ref. \cite{twistdefectchain}, where the twist-defect chain (as the edge of $\mathbb{Z}_p$ toric code) had been modeled as a conventional $\mathbb{Z}_p$ clock model. As explained in Sec. \ref{sec4.2}, the quantum clock chain realized at the edge of non-diagonal states (as well as the $\mathbb{Z}_p$ toric code) is actually the \textit{generalized} clock model, which has a much richer phase diagram for $p\geq 4$ as we discuss below.

\subsubsection{$p=4$: Ashkin-Teller model}
For $p=4$, the symmetry-enriched (self-dual) neutral sector is described by the following Hamiltonian
\begin{equation}
\begin{split}
H_{\sigma} = &-\sum_j \{J_1[\tau_j+\sigma_j\sigma^\dagger_{j-1}]\\
&+J_2[(\tau_j)^2+(\sigma_j\sigma^\dagger_{j-1})^2]+\text{H.c.}\}.
\end{split}
\end{equation}
Without loss of generality, we can assume $q=1$ (the non-trivial effect for $q>1$ would become important for $p\geq 5$). What we have got here is a one-dimensional quantum model equivalent to the highly anisotropic limit of the two-dimensional Ashkin-Teller model at self-duality \cite{AshkinTeller1943}. The corresponding phase diagram had been studied thoroughly in Ref. \cite{Kadanoff1981ATmodel}. When $J_2/J_1=0$, this model reduces to the ``conventional" $\mathbb{Z}_4$ clock model, which is equivalent to two decoupled copies of Ising models. At self-duality, the neutral sector is then gapless, characterized by the $\text{Ising}^2$ CFT which is also known as the $U(1)/\mathbb{Z}_2$ orbifold CFT at radius $R_{orb}=1$ \cite{Ginsparg88, DVVV1989}. When $J_2=J_1$, the generalized clock model has an additional $S_4$ permutation symmetry, which makes it into the four-state Potts model \cite{baxter1982exactly}. At self-duality, the neutral sector is again gapless, but this time characterized by the four-state Potts CFT, which is the $U(1)/\mathbb{Z}_2$ orbifold CFT at radius $R_{orb}=\sqrt{2}$ \cite{DVVV1989}. In fact, for $\abs{J_2/J_1}\leq 1$, there is a continuous line of criticality described by the orbifold CFT, which includes also the $\mathbb{Z}_4$ parafermion CFT \cite{Ginsparg88, Fateev:1985mm, DVVV1989}. Hence, for this region of parameter space, the $p=4$ non-diagonal state does have a gapless edge allowing for tunneling of a single electron, though the tunneling exponent is non-universal. 

Importantly, the self-dual Ashkin-Teller model is \textit{gapped} when $\abs{J_2/J_1}>1$, and this is a totally allowed region in our parameter space. Intuitively, for $J_2 \gg J_1$, the generalized clock model is dominated by the $J_2$ terms: $(\tau_j)^2$ and $(\sigma_j\sigma^\dagger_{j-1})^2$, which favor the simultaneous condensation of $\tau^2$ and $\sigma^2$ (notice that they do commute for $p=4$). This results in a partially ordered phase where $\langle \sigma^2 \rangle=\pm 1$ (there is a spontaneous symmetry breaking as either $+1$ or $-1$ is chosen) and $\langle\sigma\rangle=0$. This phase is in fact separated from a fully ordered region with $\langle \sigma \rangle \neq 0$ and a fully disordered region with $\langle\sigma^2\rangle=\langle\sigma\rangle = 0$ by two Ising transitions. For $J_2<-J_1$, the system is ordered in an antiferromagnetic frozen phase, where $\langle \sigma^2 \rangle$ equals $1$ in one sublattice and $-1$ in another. The phase diagram for the self-dual $\mathbb{Z}_4$ generalized clock chain is summarized in Fig. \ref{ATphase}. We thus conclude that, for the $p=4$ non-diagonal state, translation symmetry in the bulk (self-duality on the edge) \textit{does not} necessarily imply a gapless neutral sector on the edge. 

\begin{figure}[t!]
   \includegraphics[width=\columnwidth, height=2.8cm ]{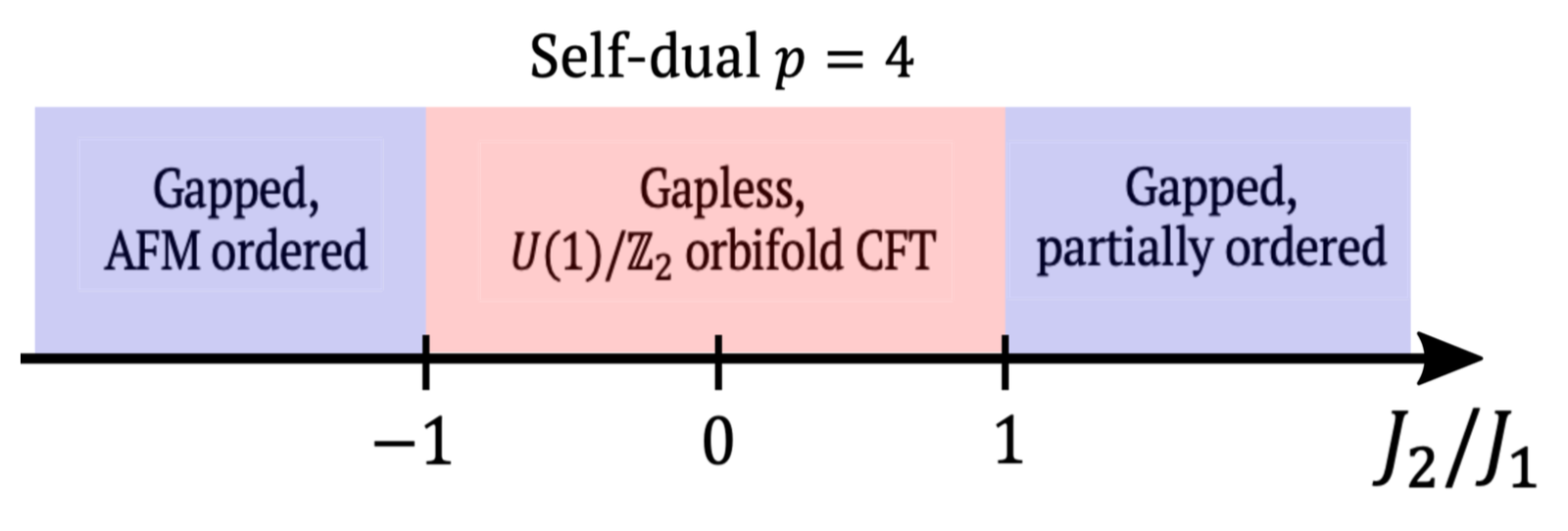}\centering
  \caption{\small{Schematic phase diagram of the self-dual $\mathbb{Z}_4$ general clock model. Notice that there exist gapped phases even at self-duality, hence there is no guarantee that the symmetric edge of the $p=4$ non-diagonal state is gapless. A detailed discussion of the complete phase diagram can be found in Ref. \cite{Kadanoff1981ATmodel}.}}
  \label{ATphase}
\end{figure}

\subsubsection{$p\geq 5$}
Similarly, for the non-diagonal states with $p\geq 5$, the neutral sector of the symmetric edge is also \textit{not} guranteed to be gapless. This is most easily demonstrated by tuning all the parameters $J_n$ to the same value, in which case the generalized clock model becomes a $p$-state Potts model. It is well-known that for $p\geq 5$ the self-dual Potts model is described by a \textit{first-order} phase transition, and is thus gapped \cite{baxter1982exactly}. In this situation, the symmetric edge would develop spontaneous dimerization appropriate for either the ordered or disordered phase. As phase coexistence could occur at a first-order transition, one may anticipate seeing both the ordered and disordered phases on the edge. Parafermion zero modes could reside at the domain walls that separate these two phases (for the general chiral model, see Ref. \cite{Fendley2012parafermionzeromode}).

On the other hand, it is interesting to ask if there can exist any gapless phase at all on the symmetric edge. The answer turns out to depend on $q$ as well. For $q = \pm1\; (\text{mod } p)$, by setting all $J_n$'s to be zero except for $J_1(=J_{p-1})$, the self-dual generalized $\mathbb{Z}_p$ model reduces to the conventional $\mathbb{Z}_p$ model at criticality, which is known to be in the gapless Berezinskii-Kosterlitz-Thouless (BKT) phase for $p\geq 5$ \cite{Dorey1996generalclock, Alcaraz1980generalclock, Ortiz2012pureclock, Ortiz2019pureclock}. However, such a gapless phase is not always allowed for a generic $q$, as can be seen by attempting (and failing) to construct a self-dual sine-Gordon representation for the BKT phase. Suppose there exists such a sine-Gordon model, then it is expected to take the following form,
\begin{equation}\label{sineGordonHam}
\begin{split}
\mathcal{H}_{SG} = &\frac{u_\sigma}{2\pi} [\widetilde{q}^2(\partial_y \phi_{\textbf{e}})^2+(\partial_y \phi_{\textbf{m}})^2]\\
&+v_{\textbf{e}} \cos(p\phi_{\textbf{e}})+ v_{\textbf{m}}\cos(p\phi_{\textbf{m}})+ ...\;,
\end{split}
\end{equation}
for some $\widetilde{q} = q\;(\text{mod }p)$. Here $\phi_{\textbf{e}}$ and $\phi_{\textbf{m}}$ are defined to satisfy
\begin{equation}\label{sineGordoncommutation}
[\phi_{\textbf{e}}(y), \phi_{\textbf{m}}(y')] =2i\pi p^{-1}H(y-y'), 
\end{equation}
where $H(y)$ is the Heaviside step function. The $\cos(p\phi_{\textbf{m}})$ term then creates vortices for $\phi_{\textbf{e}}$ with a $2\pi$-compactification, and the $\cos(p\phi_{\textbf{e}})$ term provides a $p$-state anisotropy that leads to a clock model. The clock operators can be expressed in terms of the sine-Gordon variables as follows,
\begin{equation}
e^{-i\widetilde{q}\phi_{\textbf{e}}} \sim \sigma\;\;\text{and}\;\; e^{i\phi_{\textbf{m}}} \sim \mu, 
\end{equation}
The appropriate clock algebra with $\omega=e^{2i\pi q/p}$ simply follows from the commutation relation in Eq. (\ref{sineGordoncommutation}). The duality transformation in the generalized clock model, which interchanges $\sigma \leftrightarrow \mu$, is thus equivalent to the transformation $-\tilde{q}\phi_{\textbf{e}} \leftrightarrow \phi_{\textbf{m}}$ in the sine-Gordon model. This explains the kinetic terms in Eq. (\ref{sineGordonHam}), which are chosen to ensure the self-duality. The duality would also require $v_{\textbf{m}}\cos(p\tilde{q}\phi_{\textbf{e}})$ term to appear in the Hamiltonian, but for simplicity we have swept it under the ellipsis. Notice that the two $v_{\textbf{m}}$-terms have scaling dimension $\Delta_{\textbf{m}} = p\abs{\widetilde{q}}/2 > 2$, hence they are irrelevant at low energy. 

Now a crucial observation is that the presumed dual of $\cos{p\phi_{\textbf{e}}}$ does not exist in general, because $\cos(p\phi_{\textbf{m}}/\widetilde{q})$ is not an allowed operator unless $\abs{\widetilde{q}}=1$. Without its dual, there is no term to compete with the $v_{\textbf{e}}$-term, and this would lead to gap-opening if $v_{\textbf{e}}$ flows to strong coupling. Since the scaling dimension of $\cos{p\phi_{\textbf{e}}}$ is $\Delta_{\textbf{e}}=p/(2\abs{\widetilde{q}})$, we conclude that the gapless BKT phase (or equivalently a Luttinger liquid) is allowed only when $\abs{\widetilde{q}}<p/4$, with $\widetilde{q} = q\;(\text{mod }p)$. For example, the non-diagonal state for $p=5$ and $q=1$ can have a gapless neutral sector on the symmetric edge, while for $p=5$ and $q=2$ the neutral sector can only be gapped. 

Our above discussion is not likely to be comprehensive for the symmetric edge theory of non-diagonal quantum Hall states with $p\geq 4$, and we look forward to future numerical studies that can fully characterize the phase diagram of the generalized clock chain, including the chiral model where coupling strengths are made complex. Notably, a lot have been known, either exactly or perturbatively, regarding the chiral clock/Potts model \cite{Fendley2012parafermionzeromode, Perk1989Potts, Perk1995Potts, McCoy1996Potts, Perk1997Potts}. We will leave it as a future work to examine how these known results can shape our understanding about the symmetric edge of the non-diagonal states. Nevertheless, our discussion here suffices to emphasize the distinction between the $p<4$ case and the $p\geq 4$ case: while a gapless edge is guaranteed by translation symmetry (or self-duality) in the former case, it is not guaranteed in the latter due to the possibility of having a first-order transition, and moreover, depending on the value of $q$, sometime the only possibility is to have a gapped edge that spontaneously breaks the symmetry. 

\section{\label{sec5}Summary and outlook}
In this paper, we have proposed a family of Abelian fractional quantum Hall states known as the non-diagonal states, which happen at filling fraction $\nu=p/2q$ for bosonic electrons and $\nu=p/(p+2q)$ for fermionic electrons, with $p$ and $q$ being a pair of relatively prime integers. These states are constructed using a coupled-wire model, where a single wire of Luttinger liquid is described by a non-diagonal circle CFT, and inter-wire couplings are the $pe$-tunneling. The ``non-diagonal" property dictates that a generic physical operator cannot be written as a diagonal combination of chiral and anti-chiral primary fields, which in turn strongly constrains the motion of quasiparticles in the wire construction. We realize that, in the presence of $U(1)$ charge conservation and $\mathbb{Z}$ translation symmetry of the wire model, the non-diagonal quantum Hall state possesses a non-trivial symmetry-enriched topological order. Without the translation symmetry, the non-diagonal state is identical to a strongly-clustered Laughlin state of charge-$pe$ particles, which has a $U(1)$ charge sector and a boundary characterized by the chiral Luttinger liquid. In the presence of both charge and translation symmetries, the non-diagonal state also possesses an additional neutral sector characterized by the quantum double model $\mathcal{D}(\mathbb{Z}_p)$, which has a $\mathbb{Z}_p$ topological order. Similar to Kitaev's toric code \cite{Kitaev2006exactly, Bombin2010AS} and Wen's plaquette model \cite{You2012plaquette, You2013plaquette}, the translation symmetry in the wire model acts as the $\mathbf{e}$-$\mathbf{m}$ anyonic symmetry of the $\mathbb{Z}_p$ topological order. As a result, a dislocation in the wire model, which is a termination of a wire in the bulk, acts as a twist defect for the anyonic symmetry. The non-diagonal states thus provide an electronic quantum Hall setting for realizing and testing out various ideas developed in the general theory of anyonic symmetry \cite{Teo2015twistliquid, Teo2016AS, Barkeshli2019AS}. An experimental arena for the realization of non-diagonal states maybe found in twisted materials, where an array of quasi-one-dimensional subsystems emerge with built-in translation symmetry \cite{Kennes20201dflatbands}. 

We have also investigated in detail the edge structure of non-diagonal states. For the edges perpendicular to the direction of wire, we have derived the corresponding low-energy effective Hamiltonian, which is found to consist of a chiral Luttinger liquid (for the $U(1)$ charge sector) and a generalized $p$-state quantum clock model (for the $\mathcal{D}(\mathbb{Z}_p)$ neutral sector). Discrete translation in the bulk of wire model is associated to the order-disorder duality of the clock model on the edge. For $p=2$ and $p=3$, the self-dual clock model is at a gapless critical transition, hence the non-diagonal states possess a pair of edges that are completely gapless. This is referred to as the symmetric edge, whose charge and neutral sectors are both gapless, thus allowing a single electron to be tunneled into it. In contrast, for the boundary parallel to the direction of wire, only the charge sector remains gapless and thus only allows a cluster of $p$ electrons to be tunneled into it. Hence, the non-diagonal state is anisotropic, possessing two distinct pairs of edges, as a reflection of its symmetry-enrichment. As a potential experimental probe, we have predicted the tunneling exponent for tunneling electrons from a Fermi liquid into the symmetric edge. As for $p\geq 4$, the self-dual generalized clock model on the symmetric edge acquires a richer phase diagram, which allows the neutral sector to be gapped even in the presence of symmetry. This is because the symmetric edge could be at a first-order transition, thus gapped by spontaneous symmetry breaking. It is of intellectual interest (and hopefully of practical interest in the future) to numerically study the phase diagram of the self-dual generalized $p$-clock model in greater detail, as previous studies have instead focused on the conventional clock model. We would save this for future work. 

An important future direction for us to pursue is to better characterize the non-diagonal states with the translation symmetry, equivalently the anyonic symmetry, \textit{gauged}. According to the general theory of anyonic symmetry, the gauging of anyonic symmetry in an Abelian topological phase would give rise to a non-Abelian phase \cite{Teo2015twistliquid, Teo2016AS, Barkeshli2019AS}. In the coupled-wire construction, such a gauging process concretely corresponds to the \textit{melting} of the wire model, because a dislocation (as a termination of wire) has been shown to correspond to a twist defect (\textit{i.e.} gauge flux of anyonic symmetry). Therefore, by melting the wire model of the non-diagonal anisotropic quantum Hall state, an isotropic non-Abelian quantum Hall state can be realized. We hope to develop a comprehensive theory to characterize such a state in the future. 

\textit{Note added:} The constrained motion of $\pi$-flux in weak topological superconductor (discussed here in Secs. \ref{sec2.4.2} and \ref{sec3.2.3}) is also studied recently by Rao and Sodemann \cite{Rao2020}.

\begin{acknowledgments}
The authors sincerely thank Meng Cheng, Paul Fendley, Jacques H.H. Perk and Jeffrey C.Y. Teo for helpful discussions. P.M.T. also thanks Ken K.W. Ma for discussions at the early stage of the work. This work is in part supported by the Croucher Scholarship for Doctoral Study from the Croucher Foundation (P.M.T.) and a Simons Investigator grant from the Simons Foundation (C.L.K.). 
\end{acknowledgments}

\appendix
\renewcommand{\theequation}{\thesection.\arabic{equation}}
\section{\label{secappendixa} Constructing the fermionic states}
In the main text we have explicitly constructed the non-diagonal quantum Hall states for a bosonic system. There we discussed the scattering pattern of quasiparticles in the wire model and the symmetry-enriched $\mathcal{D}(\mathbb{Z}_p)$ neutral sector that is implied. In fact, the same kind of physical phenomena appear for fermionic systems, with the electrons faithfully treated as fermions. We have alluded to this fact and also addressed the fermionic non-diagonal states in the main text, and here, we supplement with more technical details.

To account for the fermionic nature of electrons, a Jordan-Wigner string is attached to the electron operator (its bosonic version is in Eq. (\ref{phi})) to ensure anti-commutation \citep{giamarchi2003, gogolin2004}:
\begin{equation}
\psi_{\text{R}/\text{L},j} \propto e^{ \pm i (\pi\bar{\rho}x+\theta_j )} e^{i\varphi_j}
\end{equation}
To construct non-diagonal states in the coupled wire model, we adopt the same inter-wire tunneling term as depicted in Fig. \ref{setup}. More precisely, we consider the following interaction on link $\ell=j+1/2$:
\begin{equation}\label{fermiontunneling}
\begin{split}
\mathcal{V}^{(p,q)}_{\ell} &= (\psi_{\text{L},j+1}^\dagger \psi_{\text{R},j} e^{-ibx})^p \rho^q_{j+1}\rho^q_j +h.c. \\
&= e^{i(2\pi\bar{\rho}p+4\pi\bar{\rho}q-pb)x}e^{i\Theta_\ell}+h.c.\; ,
\end{split}
\end{equation}
with the link variable now defined as
\begin{equation}
\Theta_\ell = p(\varphi_j - \varphi_{j+1})+(p+2q)(\theta_j + \theta_{j+1}).
\end{equation}
As in the bosonic case, here we focus only on a coprime pair of integers $p$ and $q$. Canceling the spatial oscillatory factor in Eq. (\ref{fermiontunneling}) to guarantee momentum conservation, we obtain the filling fraction for the fermionic quantum Hall states under construction:
\begin{equation}
\nu = \frac{p}{p+2q}.
\end{equation}
The states with $p=1$ are the familiar Laughlin states, which form the diagonal series of Abelian quantum Hall states. As we show next, the $p>1$ states have interesting pattern of quasiparticle scattering that resembles the one for bosonic non-diagonal quantum Hall states, and will thus be known as fermionic non-diagonal states. There is also a non-trivial $\mathbb{Z}_p$ topological order in their neutral sector. 

The coupled wire construction proceeds in much the same way as presented in Sec. \ref{sec2.1} of the main text, which gives rise to a quantum Hall phase, with a gapped bulk where the link variables $\Theta_\ell$ are condensed at values that are integer multiples of $2\pi$, and with a pair of gapless chiral edges now described by a circle CFT at radius $R=\sqrt{p/(p+2q)}$ . Here we note that the decoupled chiral bosonic modes in a single wire, originally defined for a bosonic system according to Eq. (\ref{chiralboson}), are now modified to account for the Jordan-Wigner string of the electron:
\begin{subequations}\label{fermionchiralboson}
\begin{align}
\phi^\text{R}_j &= p\varphi_j + (p+2q)\theta_j, \\
\phi^\text{L}_j &= p\varphi_j - (p+2q)\theta_j,  
\end{align}
\end{subequations}
and their commutation relations become
\begin{equation}
[\partial_x\phi^{\tilde{r}}_j(x),\phi^{\tilde{r}'}_{j'} (x')] = 2i\pi p (p+2q) \tilde{r}\delta_{\tilde{r}\tilde{r}'}\delta_{jj'}\delta(x-x'),
\end{equation}
where $\tilde{r},\tilde{r}'=\text{R}/\text{L}=+1/-1$. Up to this point, it should be clear that many changes from the bosonic case to the fermionic case can be accounted for by simply taking $2q  \mapsto p+2q$. Quasiparticle excitations in the wire model again correspond to $2\pi$-kinks in the link variables $\Theta_\ell$. Analogous to Eq. (\ref{QPannihilation}), the annihilation operator for the minimal quasiparticle is expressed as
\begin{equation}
\Psi^{\text{R}/\text{L}}_{e/(p+2q), \ell} = e^{\frac{i}{p(p+2q)}\phi^{\text{R}/\text{L}}_{j/j+1}},
\end{equation}
where the charge of the minimal quasiparticle in the fermionic phase is $e/(p+2q)$. A charge-$pe$ excitation can be created/annihilated by the physical operator $e^{i\phi^{\text{R}/\text{L}}}$, and is thus treated as the trivial quasiparticle that is identified with the vacuum. Hence there are $N=p(p+2q)$ distinct Abelian excitations within each link. 

Notice that, unlike in the bosonic case, here we have to carefully distinguish the term ``local" from the term ``physical". Since our system is made up of electrons, we would refer to an operator that can be expressed as a product of electronic operators as ``physical", which are allowed to appear in the Hamiltonian. Since the fermionic electron is strictly speaking non-local, local operators form only a subset of physical operators, which do not change the fermion-parity. For certain non-diagonal fermionic states, the fermion-parity symmetry can replace the role of charge conservation in constraining the motion of quasiparticles. This particular distinction between bosonic and fermionic states is discussed in Sec. \ref{sec3.2}.

A generic physical scattering operator (in the above sense) can be expressed as 
\begin{equation}\label{fermioniclocalop}
\mathcal{O}^{\{r,s\}}_j = e^{i[r\varphi_j-(r+2s)\theta_j]}
\end{equation}
with $r,s \in \mathbb{Z}$. To interpret its effect of scattering quasiparticles, we make a change of variables to the chiral bosonic fields and obtain
\begin{equation}\label{fermioniclocalop2}
\mathcal{O}^{\{r,s\}}_j = \text{exp} \frac{i}{p(p+2q)}[(qr-ps)\phi^\text{R}_j+(qr+ps+pr)\phi^\text{L}_j].
\end{equation}
This is telling us that $\mathcal{O}^{(r,s)}_j$ would scatter a quasiparticle of charge $e(qr-ps)/(p+2q)$ residing on link $j+1/2$ to another quasiparticle of charge $-e(qr+ps+pr)/(p+2q)$ residing on link $j-1/2$. For the particular cases with $(r,s) = \pm(p,q), \pm(-p,p+q)$, the operator is either creating or annihilating a trivial quasiparticle of charge $pe$. For systematic analysis of quasiparticle scattering, we can organize the operators onto a lattice, similar to the bosonic case in Fig. \ref{localoperator}. The difference is that the points corresponding to physical operators are now ordered in a checker-board pattern, because of the additional Jordan-Wigner string $e^{ir\theta}$ in Eq. (\ref{fermioniclocalop}). Two representative scenarios for the fermionic case are demonstrated in Fig. \ref{flocaloperator}. 

It is again important to distinguish the scattering operators that are \textit{charged} (with $r\neq 0$) from those that are \textit{charge-neutral} (with $ r=0$). Quasiparticles that are scattered by the charge-neutral operators are the only quasiparticles that can be scattered across a single wire under the constraint of locality and charge conservation. From Eq. (\ref{fermioniclocalop2}), it is clear that these quasiparticles carry charge of integral multiples of $pe/(p+2q)$. For states with $p=1$, these are all the quasiparticles. For states with $p>1$, there exist non-trivial quasiparticles, including the minimal quasiparticle of charge $e/(p+2q)$, that cannot be scattered across only a \textit{single} wire. Instead, they have to hop across \textit{two} wires at a time through the following local operator which preserves charge:
\begin{equation}\label{fscatter2wire}
\mathcal{O}^{\{-r, r+s\}}_{j-1} \mathcal{O}^{\{r,s\}}_j \propto \text{exp } i[\frac{(qr-ps)}{p(p+2q)}(\phi^\text{R}_j-\phi^\text{L}_{j-1})]. 
\end{equation}
Since $p$ and $q$ are assumed to be relatively prime, the B\'ezout's lemma  guarantees the existence of integral solutions $(r,s)$ such that $qr-ps=1$. Hence the minimal quasiparticle in a non-diagonal state, though cannot be scattered across a single wire, can indeed be scattered across two at a time. This defining feature of non-diagonal quantum Hall states in the coupled wire model would distinguish quasiparticles on the \textit{even} links from those on the \textit{odd} links, and eventually reveal a hidden $\mathbb{Z}_p$ toric code in the neutral sector.

\bibliographystyle{apsrev4-1.bst}
\bibliography{ToricCodeQH}

\begin{thebibliography}{99}
\bibitem [{Tdu()}]{Tduality}%
  \BibitemOpen
  \href@noop {} {}\bibinfo {note} {Notice that the radius $R$ used here, which
  is consistent with that defined in \citep{Ginsparg88}, is related to the
  radius $R_{\rm BYB}$ defined in \citep{BYB} by the relation $R_{\rm
  BYB}=2R$.}\BibitemShut {Stop}%
\bibitem [{clo()}]{clockfermionic}%
  \BibitemOpen
  \href@noop {} {}\bibinfo {note} {One would substitute $2q\mapsto p+2q$ in
  order to understand the non-diagonal fermionic states. In this case, $\omega
  = e^{2i\pi q/p} \mapsto -\omega$. One should note that the extra minus sign
  would be compensated by including Klein factors in the electron operators so
  that anti-commutation in the $y$-direction is treated properly. At the end,
  the neutral sector is also a $p$-state clock model for the fermionic
  states.}\BibitemShut {Stop}%
\end{thebibliography}

\end{document}